\documentclass[fleqn,usenatbib]{mnras}
\usepackage{newtxtext,newtxmath}
\usepackage[T1]{fontenc}

\DeclareRobustCommand{\VAN}[3]{#2}
\let\VANthebibliography\thebibliography
\def\thebibliography{\DeclareRobustCommand{\VAN}[3]{##3}\VANthebibliography}

\usepackage{graphicx}	
\usepackage{amsmath}	
\usepackage{CJK}

\usepackage{ulem}

\newcommand{\ys}{\bgroup\markoverwith{\textcolor[rgb]{1.0, 0, 1.0}{\rule[0.5ex]{8pt}{1.5pt}}}\ULon}

\title[Mean-motion Resonances in Turbulent Discs]{Capture and Escape of Planetary Mean-motion Resonances in Turbulent Discs}

\author[Chen \& Wu et al. 2025]{Yi-Xian Chen (陈逸贤),$^{2}$\footnote[1]~
Yinhao Wu (吴寅昊),$^{1}$\thanks{Co-first authors with equal contributions.}
Ya-Ping Li (李亚平),$^{3}$~
Douglas N. C. Lin (林潮)$^{4,5},$~
\newauthor 
Richard Alexander$^{1},$~
Sergei Nayakshin$^{1},$~
Fei Dai (戴飞)$^{6}$~
\\
$^{1}$School of Physics and Astronomy, University of Leicester, Leicester LE1 7RH, UK \href{mailto:email@domain}{(yw505@leicester.ac.uk)}\\
$^{2}$Department of Astrophysical Sciences, Princeton University, 4 Ivy Lane, Princeton, NJ 08544, USA \href{mailto:email@domain}{(yc9993@princeton.edu)}\\
$^{3}$Shanghai Astronomical Observatory, Chinese Academy of Sciences, Shanghai 200030, China \href{mailto:email@domain}{(liyp@shao.ac.cn)}\\
$^{4}$Department of Astronomy \& Astrophysics, University of California, Santa Cruz, CA 95064, USA\\
$^{5}$Institute for Advanced Studies, Tsinghua University, Beijing 100084, China\\
$^{6}$Institute for Astronomy, University of Hawai'i, 2680 Woodlawn Drive, Honolulu, HI 96822, USA
}

\date{Accepted XXX. Received YYY; in original form ZZZ}

\pubyear{\the\year{}}

\begin{document}
\begin{CJK*}{UTF8}{gbsn} 
\label{firstpage}
\pagerange{\pageref{firstpage}--\pageref{lastpage}}
\maketitle

\begin{abstract}
Mean-motion resonances (MMRs) form through convergent disc migration of planet pairs, 
which may be disrupted by dynamical instabilities after protoplanetary disc (PPD) dispersal. 
This scenario is supported by recent analysis of TESS data showing that neighboring planet pairs in younger planetary systems are {closer to resonance}. 
To study stability of MMRs during migration, 
we perform hydrodynamical simulations of migrating planet pairs in PPDs, 
comparing the effect of laminar viscosity and realistic turbulence. 
We find stable 3:2 resonance capture for terrestrial planet pairs migrating in a moderately massive PPD,  
insensitive to a range of laminar viscosity ($\alpha = 10^{-3} - 10^{-1}$).
However, realistic turbulence enhances overstability by sustaining higher equilibrium eccentricities and a positive growth rate in libration amplitude, 
ultimately leading to resonance escape. 
The equilibrium eccentricity growth rates decrease as planets migrate into tighter and more stable 4:3 and 5:4 MMRs. 
Our results suggest that active disc turbulence broadens the parameter space for overstability, 
causing planet pairs to end up in closer-in orbital separations. Libration within MMR typically 
lead to deviation from exact period ratio $|\Delta| \sim 0.5\%$, 
which alone is insufficient to produce the typical dispersion of $|\Delta| \sim 1-3\%$ in TESS data, suggesting that post-migration dynamical processes are needed to further amplify the offset.
\end{abstract}




\section{Introduction}

Orbital migration 
plays a crucial role in planet formation, 
ultimately shaping 
the architecture of planetary systems. 
Disc-driven migration occurs in the early stages of protoplanetary disc (PPD) evolution, 
when planets tidally interact with their natal discs. 
Type I migration primarily affects low-mass planets \citep{Goldreich1980,Ward1997,Tanaka2002}, 
typically up to a few Earth masses around Sun-like stars, 
before they reach the thermal mass to open up a gap in the disc \citep{LinPapaloizou1986}. 
This process influences the formation of super-Earths and proto-gas-giant cores \citep{ida2008,benz2014,liu2015}. 
Migration torques, 
set by the disc structure \citep{Paardekooper2010,Kley2012,Paardekooper2023}, 
usually leads to inward motion on a timescale shorter than the typical lifetime of PPDs. Recent studies have also found that additional physical processes, such as disc winds \citep{McNally2017,McNally2018,Wu-Chen2025}, dust feedback \citep{Guilera2023,Hou2024,Hou2025}, and background drift \citep{Wu2024BDHI}, may also have the potential to modify the torques exerted on low-mass planets.

Mean-motion Resonances (MMRs) or 
resonant chains naturally arise from convergent migration of planet pairs \citep{LeePeale2002} 
particularly when migration is relatively slow, 
though the exact conditions of resonant capture
remain an active area of research 
\citep{Goldreich-Schlichting2014, HuangOrmel2023, Batygin2023, Wong2024, Lin2025, Keller2025}. 
After PPD dispersal, 
these chains are expected to gradually break due to dynamical instabilities \citep{Deck2013, Izidoro2017, Lammers2024}, 
leading to late-stage compact systems that are typically not in resonance.
The convergent migration scenario is supported by analysis of TESS data in \citet{Dai2024}, 
who show that across an age spectrum, 
earlier planetary systems 
are {more likely to be near-resonant}.

Notably, 
these studies also find that 
observed period ratios often deviate from perfect commensurability by a fraction of $\Delta \sim$ from $-1\%$ to $3\%$, 
a feature also present in the Kepler catalogue \citep{Fabrycky2014}.
This is commonly attributed to tidal dissipation after disc dispersal \citep{PapaloizouTerquem2010, LithwickWu2012, Lee2013}, though it has recently been suggested that the offset from the perfect period
commensurability with a few percent could be attributed to the interference density wave when the planet pair is close to MMR \citep{Yangli2024}. 
However, 
the persistence of this $\Delta$ feature even in very young planetary systems, 
as emphasized by \citet{Dai2024}, 
suggests that at least some fraction of this deviation may originate from migration itself. This is because producing such 
$\Delta$ by dissipation alone would require an 
unrealistically low total quality factor $Q$ (a measurement of material dissipation) for young planets \citep{Xu-Dai2025}. 

This leads us to explore the possibility that planets may not be trapped 
in exact MMRs during convergent migration 
in the first place.
The theory of 
resonant overstability, developed in the study of Saturnian satellites \citep{MeyerWisdom2008}, 
has been invoked by \citet{Goldreich-Schlichting2014} as a potential explanation of non-negligible $\Delta$ during migration. 
Namely, 
when
planet pairs migrate into MMRs, 
small perturbation in their eccentricities and period ratio evolves with time as $\propto \exp(s + i\omega)t$. 
Both the libration frequency $\omega$ 
and the growth rate $s$ are explicit functions of the equilibrium eccentricity $e_{\rm eq}$, 
which itself depends on the planet masses $M_1, M_2$ and disc aspect ratio $h$. 
Analytical theory predicts that these oscillations are generally damped out ($s<0$) for planet-to-star mass ratio $q = M/M_\star \gtrsim h^3$, while $s>0$ for lower mass $q \lesssim h^3$, 
the latter leads to either (i)
saturation towards an equilibrium libration amplitude, or 
(ii) escape out of resonance within an eccentricity damping timescale, and subsequent migration towards the next (closer-in) 
resonance. 
Disc dispersal during these librations will imprint such amplitudes into the current $\Delta$ distribution, 
even before the occurence of post-disc dynamical instabilities. 
This dichotomy is confirmed by 
recent hydrodynamic simulations of planet pairs migrating in and out of 2:1 resonances \citep{Hands-Alexander2018,Ataiee-Kley2021,Afkanpour2024MMR}, 
although it's worth noting that
deviations from analytical theory can arise in low-viscosity cases due to partial gap-opening effects. 

Nevertheless, 
as in classic simulation of planet migration in PPDs, 
turbulence
is modeled as a laminar viscosity term
that's expected to 
solely enhance dissipation, 
damp waves and interference, 
and prevent gap-opening \citep[e.g.][]{Masset2000,Paardekooper2010, Kanagawa2015}. 
Realistic turbulence from 
magneto-rotational instability (MRI) \citep{Beckwith2011,Simon2012,Rea-etal.2024} and/or gravitational instability (GI)  \citep{Rice2003,Deng2017} 
present in PPDs, 
on the other hand, 
introduce trans-sonic scale turbulent eddies. 
These eddies could generate stochastic torques \citep{Nelson2005,Wu2024chaotic,Kubli2025} 
and excite oscillation from exact resonances, 
potentially influencing the equilibrium eccentricity $e_{\rm eq}$ 
as well as the growth rate $s$, 
which differs from laminar cases. The 
realistic effect of 
such turbulence on 
resonance capture remains 
to be studied by 
hydrodynamic simulations incorporating active turbulent prescriptions.

This paper is organized as follows: 
We introduce our numerical setup of modeling planet-disc interaction in \S \ref{sec:setup} following \citet{BaruteauLin2010,ChenLin2023, Wu2024chaotic}. 
In \S \ref{sec:results}, 
we present results from our simulations, 
focusing on the comparison between 
laminar viscosity versus active turbulence. 
In \S \ref{sec:analysis}, 
we analyze equilibrium eccentricities and growth rates of perturbation at MMRs seen in our simulations 
under the \citet{Goldreich-Schlichting2014} framework and conclude that
active turbulence enhances both $e_{\rm eq}$ and $s$, favoring overstability. 
We discuss the implication of
our findings and lay out 
future prospects in \S \ref{sec:summary}.

\section{Numerical Setup}
\label{sec:setup}

To explore multi-planet migration in PPDs, 
we use the 
hydrodynamic grid-based code \texttt{FARGO3D} \citep{FARGO3D}, 
which is the successor of the \texttt{FARGO} code \citep{Masset2000}. 

For our disc model, 
we set a locally isothermal disc with an aspect ratio $h \equiv H/r= h_0$, 
here the reference aspect ratio is $h_0 = 0.03$, consistent with most PPD models at close-in regions $< 1$au \citep{GaraudLin2007,Chiang2010,Chiang2013,Chen2020}. 
We also set the initial disc gas viscosity to be $\nu = \alpha H^2 \Omega$ where $\alpha$ is 
a dimensionless parameter\citep{shakura1973} and 
surface density profile to be $\Sigma = \Sigma_0 (r/r_0)^{-1/2} \left[M_{\star}/r_{0}^{2}\right]$, 
where $\Sigma_0$ is a normalization constant relevant to disc mass, $r_0$ is the code unit length and the initial inner planet's orbital radius,
$M_{\star}$ is the mass of the host star and the code unit for mass
and $\Omega=\sqrt{G M_\star/r^3}$ is the Keplerian angular frequency.  
Because of the low disc masses considered in our simulations,
we neglect self-gravity of the disc.  The indirect term associated with the stellar motion is enabled (i.e., using the \texttt{GASINDIRECTTERM} option),
as the star is fixed at the origin of the reference frame.

To model active turbulence for comparison against laminar $\alpha$, 
we modify the code with a 
phenomenological turbulence prescription \citep{Laughlin2004,BaruteauLin2010,ChenLin2023,Wu2024chaotic} to study planet migration in turbulent discs. 
We add a fluctuating potential $\Phi_{\rm turb}$ to the momentum equation consisted of 50 stochastic modes at each timestep:
\begin{equation}
    \Phi_{\rm turb }(r, \phi, t)=\gamma r^{2} \Omega^{2} \sum_{k=1}^{50} \Lambda_{k}(m_k, r, \varphi, t),
    \label{eqn:phiturb}
\end{equation}
where $\gamma$ is a dimensionless characteristic amplitude of turbulence. Each mode, denoted as $\Lambda_{k}$, 
is given by
\begin{equation}\label{Lambda}
    \Lambda_{k} = \xi_k e^{-(r-r_k)^2/\sigma_k^2} \cos(m_k\phi -\phi_k-\Omega_k \tilde{t}_k) \sin (\pi \tilde{t}_k/\Delta t_k),
\end{equation}
which is associated with 
a wavenumber $m_k$ drawn from a logarithmically uniform distribution between $m=1$ and the maximum value $m_{\rm max}$ corresponding to azimuthal grid scale.
The initial radial position $r_k$ and azimuthal angle $\phi_k$ of each mode are selected from a uniform distribution. 
The radial extent of each mode is $\sigma_k = \pi r_k /4m$. 
Modes activate at time $t_{0,k}$ 
and last for $\Delta t_k = 0.2\pi r_k / m c_s$, 
where $c_s$ denotes the local sound speed. 
$\Omega_k$ represents the Keplerian frequency at $r_k$, $\tilde{t}_k = t - t_{0,k}$, 
and $\xi_k$ is a dimensionless constant sampled from a Gaussian distribution with unit width. 
Following \citet{BaruteauLin2010}, 
the parameter choices for this turbulence driver 
emulate a Kolmogorov cascade power spectrum, 
maintaining a $m^{-5/3}$ scaling law.

The relationship between $\gamma$ and the time-average effective Reynolds stress parameter $\langle \alpha_R \rangle$ (hereafter referred to as $\langle \alpha \rangle$) 
generated by this turbulence driver is calibrated to be \citep{BaruteauLin2010,ChenLin2023,Wu2024chaotic}:
\begin{equation}\label{alpha_R}
    \langle \alpha \rangle \simeq 35 (\gamma/h_0)^2.
\end{equation}  

{Although recent observational studies of line broadening have constrained the turbulence strength in nearby PPDs at large radial distances ($>30$ au) to be $\alpha<10^{-4}-10^{-3}$ \citep[e.g.,][]{Flaherty2020,Lesur-etal.2023-PPVII, Rosotti2023}, 
turbulence level in the inner disc regions ($<1$ au) 
remains unclear due to limitations of current measurement techniques \citep{Salyk2008,Romero-Mirza2024} and the possibility that existing observations may not be sensitive enough to detect viscous spreading 
\citep{Alexander-etal.2023}. Resolved observation Yound Stellar Object SVS 13 suggest that $\alpha\sim10^{-2}$ at $\lesssim 0.3$au \citep{Carr-etal.2004}. 
Moreover, recent studies indicate that certain FU Ori-type outbursts can be explained by $\alpha\sim0.03$ 
\citep{Cleaver-etal.2023,Nayakshin-etal.2024}, 
aligning with what we expect from theories of MRI when the temperature is high enough or when $\Sigma$ is low enough to be penetrated by cosmic rays \citep{Gammie1996}. 
In this study, 
we explore the range of $ \langle \alpha \rangle $ between $10^{-3}$ to $10^{-1}$. }

Our simulations are performed in a 2D $\left(r,\phi\right)$ coordinate system, 
with a computational domain ranging from 0.4 $r_0$ to 3.2 $r_0$ radially and 0 to $2\pi$ azimuthally. 
The domain is resolved by 512 logarithmic grid cells in the radial direction and 1536 grid cells in the azimuthal direction. 
We apply wave-damping radial boundary conditions \citep{deValBorro2006}, 
activating the \texttt{KEPLERIAN2DDENS} and \texttt{STOCKHOLM options} in \texttt{FARGO3D}. 
This setup is same as the \texttt{Evanescent} boundary conditions applied in \cite{BaruteauLin2010} and \cite{Wu2024chaotic}. 
To avoid numerical issues caused by turbulence damping at the boundaries, 
we did not extend the turbulence across the 
entire disc but limit it to a radial range from 0.5 $r_0$ to 2.8 $r_0$. 

In all simulations presented in this work, 
the planet system setup
consists of Planet 1 (P1) initially positioned at an orbital radius of $r_{1} = 1.0~r_0$ 
with a planet-to-stellar mass ratio of $q_{1} = 5 \times 10^{-6}$, 
and Planet 2 (P2) initially positioned at an orbital radius of $r_{2} = 1.4~r_0$ (for laminar cases) 
or $1.35~r_0$ (for turbulent cases) with a planet-to-stellar mass ratio of $q_{2} = 1.5 \times 10^{-5}$. 
Both orbital separations lie just beyond the 3:2 MMR. 
The mass ratios correspond to $M_1 = 1.7 M_\oplus, M_2 = 5 M_\oplus$ for a solar mass host star $M_\star = M_\odot$.
We fully account for both planet-disc interaction and planet-planet interaction, 
allowing the planets to migrate freely. 

As an indicator for MMR, during our runs we measure the inner planet's (P1's) resonant angle close to each $j:j+1$ order MMR as:
\begin{equation}
    Q = (j + 1)\lambda_2 - j \lambda_1 -\varpi_1,
\end{equation}
where $\lambda_1$ and $\varpi_1$ are mean longitude and longitude of pericenter of P1, and $\lambda_2$ is mean longitude of P2 \citep{Murray-Dermott1999}. 
For a stable resonance, 
$Q$ is expected to converge to some value close to $0$ or $\pi$ \citep{Goldreich-Schlichting2014,Terquem2019}. 

For each resonance, 
we generally consider planets to be trapped in resonance at $t_{\rm in}$ when the time evolution of period ratio $P_2/P_1$ flattens (a distinct transition, see \S \ref{sec:results}). 
In some turbulent cases, 
we also consider planets to migrate out of resonance at $t_{\rm out}$ when $P_2/P_1$ deviates from commensurability by  $>1\%$. 

We measure the average equilibrium eccentricity $e_{\rm  eq}$ for each resonance from $t_{\rm in}$ to  $t_{ \rm out}$ (or the end of the simulation, if we do not see resonance escape).





\section{Results}
\label{sec:results}

\subsection{Multi Planets Migration in Laminar Discs}\label{sec:3.1}

\begin{figure}
\centering
\includegraphics[width=0.99\hsize]{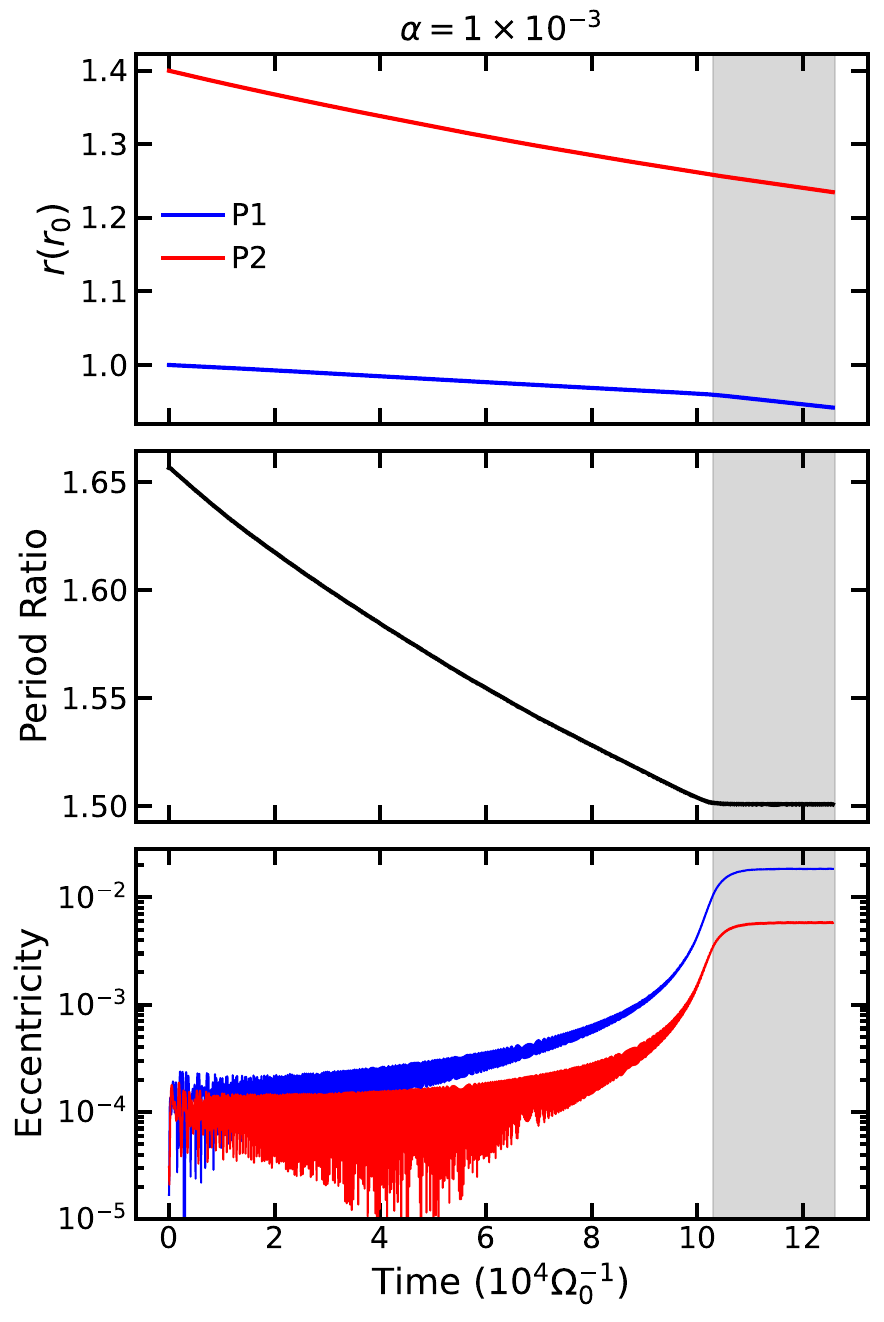}
\caption{The three panels illustrate simulation results for our fiducial case (with $\alpha=1\times10^{-3}$ and $\Sigma_0=2\times10^{-5}$, as shown in the title). 
From top to bottom panel, we show time evolution of different orbital parameters: the radial positions of P1 (blue) and P2 (red); the period ratio between P1 and P2; 
the eccentricities of P1 and P2. 
The time axis is expressed in units of the dynamical timescale at P1's initial position (i.e., $r_p = 1r_0$). 
The two planets enters a 3:2 resonance state after approximately $t_{\rm in} = 1.03\times 10^5\Omega_0^{-1}$ remained stable until the end of the simulation. Gray bands indicate the time ``in resonance" during which we average $\langle e_1\rangle $ and $\langle e_2\rangle$.}\label{fig:fiducial}
\end{figure}

\begin{figure}
\centering
\includegraphics[width=1\hsize]{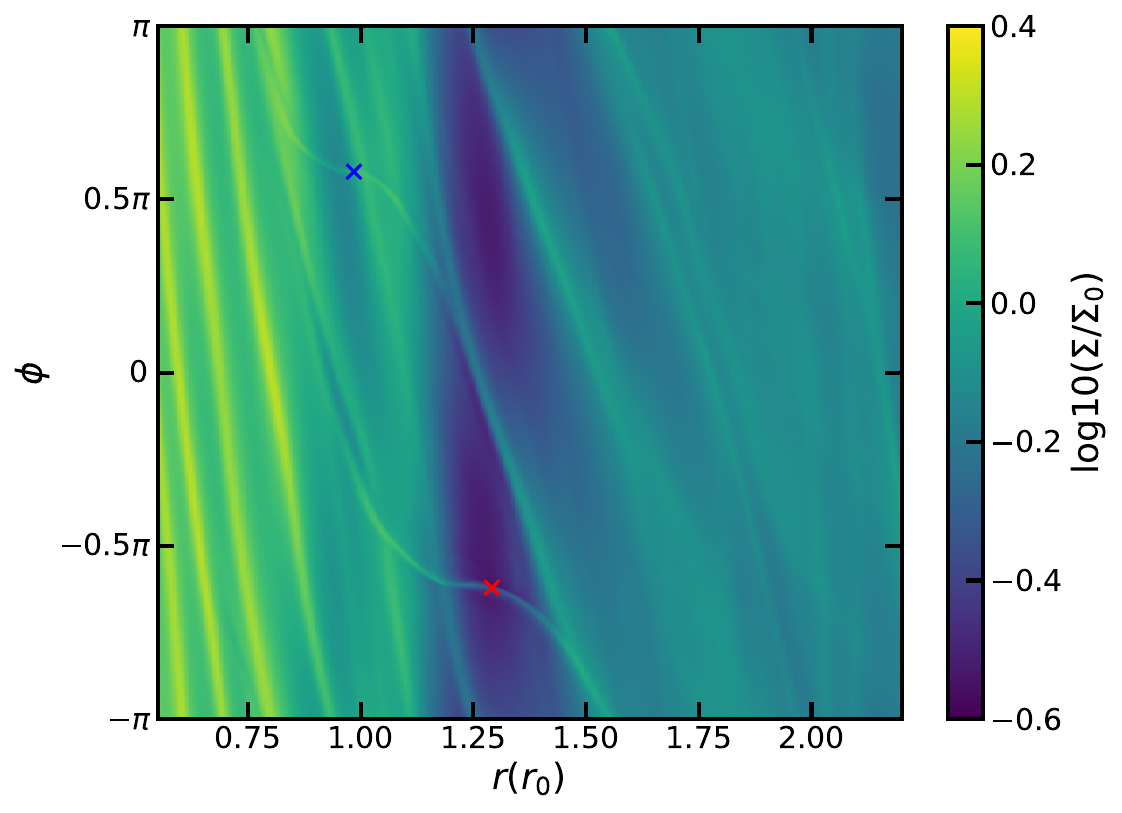}
\caption{Typical surface density distribution for the fiducial case near the end of the simulation, where the planet pairs have captured into a 3:2 MMR. We use blue and red "x" symbols to indicate the positions of the inner (P1) and outer (P2) planets, respectively.}\label{fig:snapshot}
\end{figure}

\begin{figure}
\centering
\includegraphics[width=0.85\hsize]{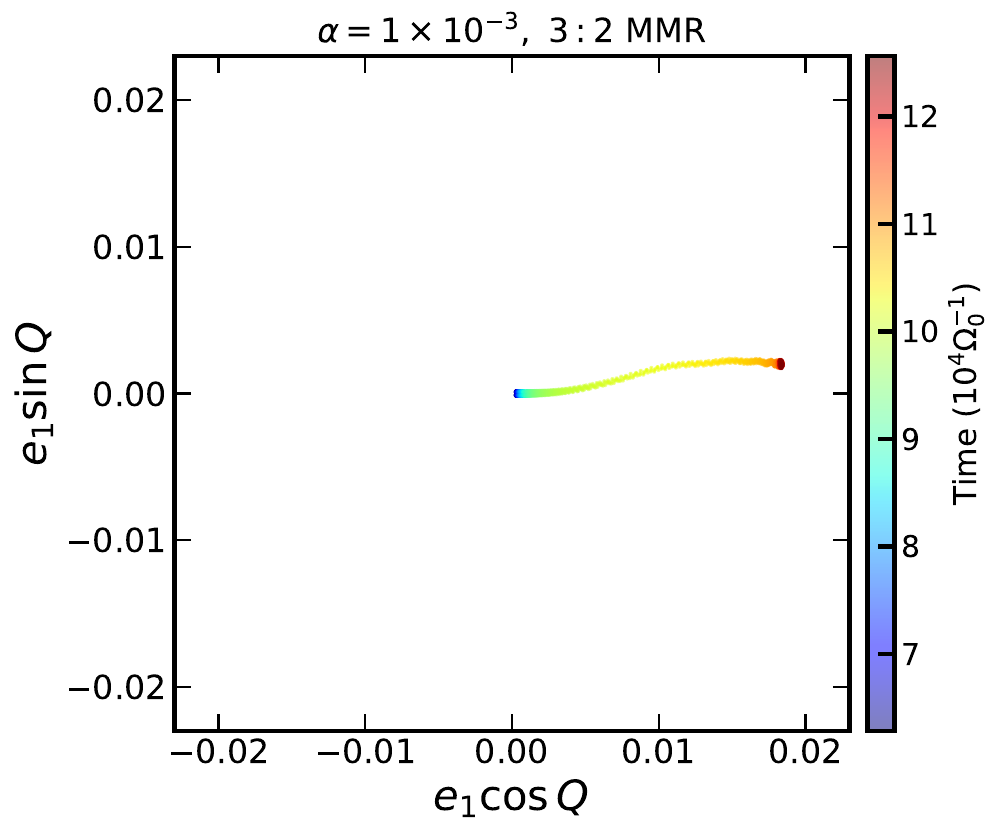}
\caption{The time evolution of the resonant angle phase space of the inner planet for our fiducial model (laminar case with $\alpha=1\times10^{-3}$, see in $\S$\ref{sec:3.1}) as the planet pairs migrate into the 3:2 resonance. 
Colour represents time. 
}\label{fig:alpha_1e-3_Qphase}
\end{figure}

We first present results from the fiducial laminar simulation with $\Sigma_0 = 2\times 10^{-5}$ and $\alpha = 10^{-3}$. 
We show in Figure \ref{fig:fiducial} 
the time evolution of a series of orbital parameters (from top to bottom: 
the semi-major axis, 
the period ratio, and the eccentricities $e$) obtained from our fiducial case. 
Using the dynamical timescale $\Omega_0^{-1}$
at P1's initial position $r_p=1.0 r_0$ as the reference time unit (which naturally arises from our choice of $G = M_\star = r_0$ as code unit), 
the planet pair enters a 3:2 MMR state after undergoing approximately $t_{\rm in} = 1.03\times 10^5 \Omega_0^{-1} \approx 1.65\times 10^4 $  orbits of Type I migration. 
The system remains trapped in this resonance until the end of the simulation. 
During resonance,  
the planets' eccentricities increase significantly by approximately two orders of magnitude, 
with the inner planet (P1) reaching as high as $\langle e_1 \rangle = 0.018$ and the outer planet (P2) $\langle e_2 \rangle = 0.006$. 
Note we have highlighted the time interval for performing time-averages of eccentricities from $t_{\rm in}$ to the end of simulation with gray shading in Figure \ref{fig:fiducial}. 
Figure \ref{fig:snapshot} shows a typical surface density distribution snapshot during the 3:2 MMR in this simulation, 
where P2 causes a partial gap but gas depletion around the orbit of P1 is not significant. There are signs of interference between density waves generated by the planet pair.
In Figure \ref{fig:alpha_1e-3_Qphase}, 
we present the phase diagram of the time evolution of P1's resonance angle for the fiducial case (with P1's eccentricity as radial coordinate). 
It can be seen that the system remains at a fixed point for the resonance angle and eccentricity in the end of simulation without showing signs of overstability.
The eccentricity $e_1$ converges to $\langle e_1 \rangle$ and $\sin Q$ towards a small value $\ll 1$ \citep{Goldreich-Schlichting2014}. 
We have confirmed that the system is still trapped in the same fixed point when it evolves by further $3\times10^{4}\Omega_{0}^{-1}$ without showing substantial change.

\begin{figure}
\centering
\includegraphics[width=1\hsize]{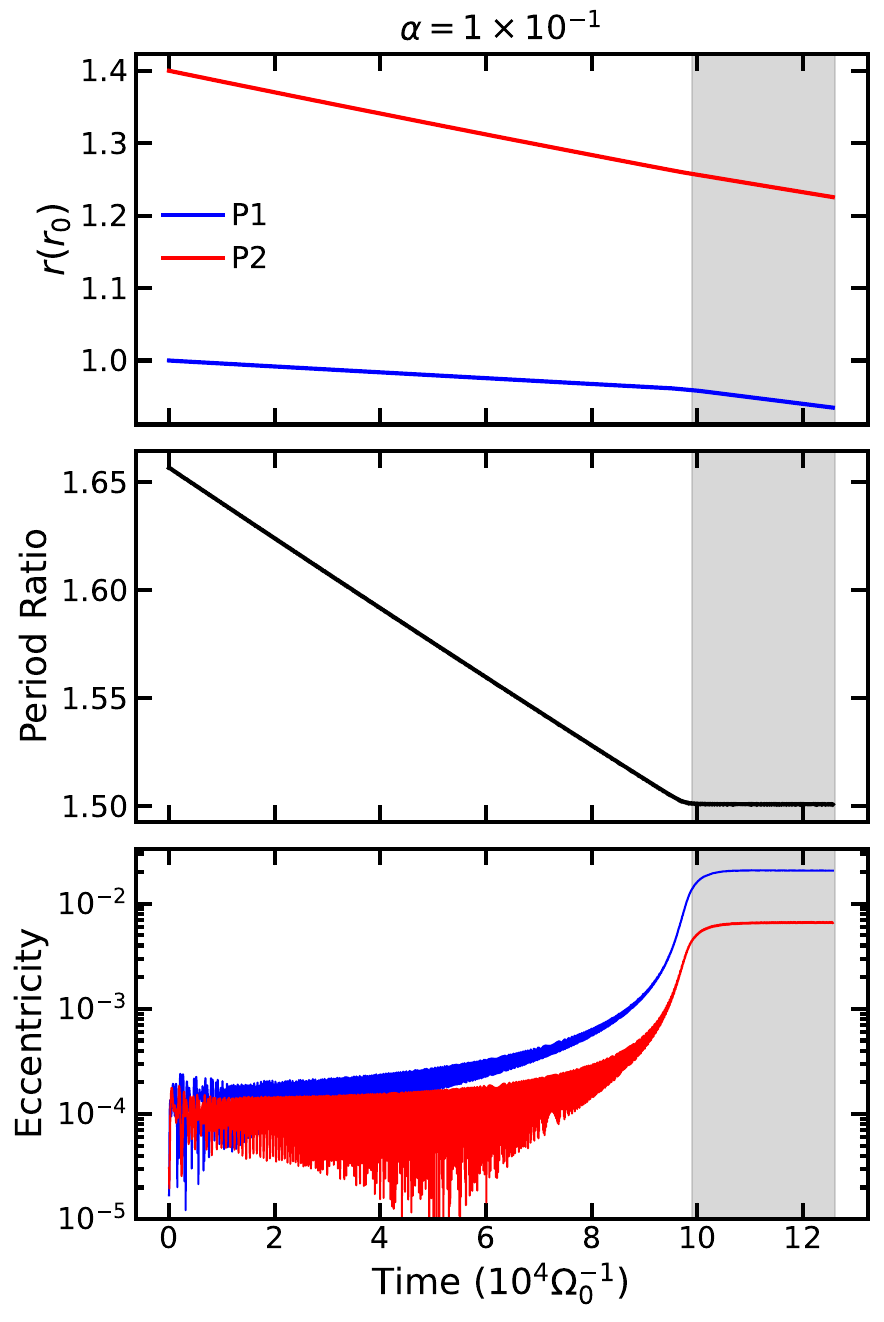}
\caption{Similar to Figure \ref{fig:fiducial}, but for strong viscosity case (with $\alpha=1\times10^{-1}$ and $\Sigma_0=2\times10^{-5}$). 
The two planets entered a 3:2 resonance state after approximately $t_{\rm in} = 9.7\times 10^4\Omega_0^{-1}$ 
and remained stable until the end of the simulation.}
\label{fig:alpha1e-1}
\end{figure}

We also present the results of orbital parameters under 
higher laminar viscosity ($\alpha=1\times10^{-1}$) in Figure \ref{fig:alpha1e-1}. 
Overall, planets migrate slightly faster and the 3:2 MMR is reached earlier by $\sim 4000 \Omega_0^{-1} \approx 600$ orbits compared to the $\alpha=1\times10^{-3}$ case.  
Aside from this, 
the evolution of other orbital parameters closely resembles 
that of the fiducial simulation, 
confirming that equilibrium eccentricity and overstability in MMRs
remain unaffected by viscosity as long as gap-opening is not significant and migration 
stays in the Type I regime 
\citep{Afkanpour2024MMR}.

It is worth noting that our fiducial choice of $\Sigma_0$ guarantees 
a mild initial migration rate, 
allowing the planets to gradually become trapped in the 3:2 MMR. 
We show in Appendix \ref{appendix: disc-model} that planetary systems with more massive discs exhibit migration rates enhanced by approximately a factor of $\sim 5$, driving planets to bypass the 3:2 resonance configuration, 
consistent with theoretical expectations \citep{Lin2025}.
Nevertheless, 
whether resonant planets escape MMR due to overstability is a
separate issue 
independent of the capture criterion, 
the latter of which lies beyond the main scope of our paper.

\begin{figure}
\centering
\includegraphics[width=1\hsize]{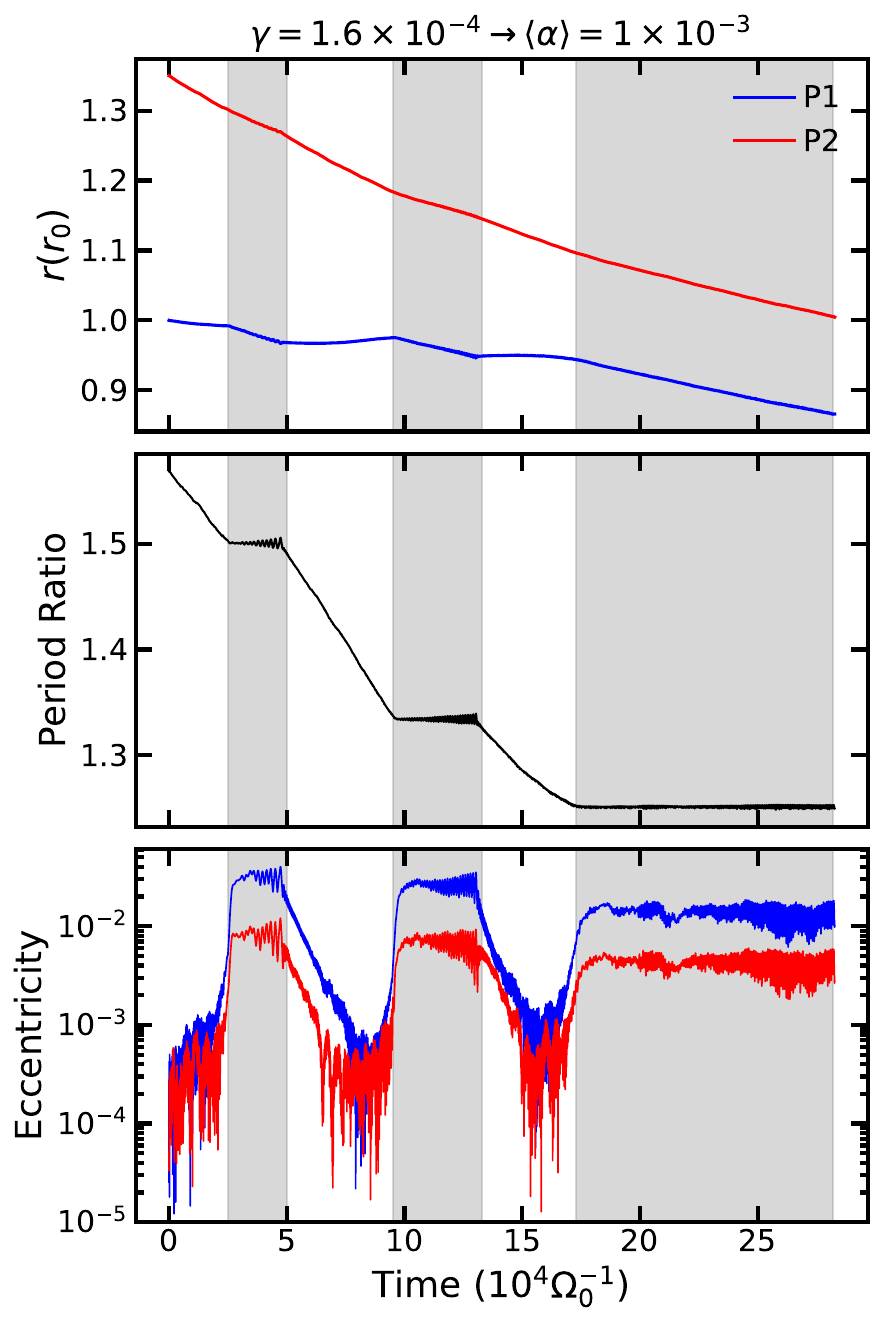}
\caption{Similar to Figure \ref{fig:fiducial}, but for 
weak turbulent case (with $\gamma=1.6\times10^{-4}$, corresponding to an effective viscosity stress parameter $\langle \alpha\rangle =1\times10^{3}$). 
The planet pair traversed a total of three MMR states: 3:2, 4:3, and 5:4 and experienced two instances of resonance breaking.}\label{fig:gamma1e-4}
\end{figure}

\begin{figure}
\centering
\includegraphics[width=1\hsize]{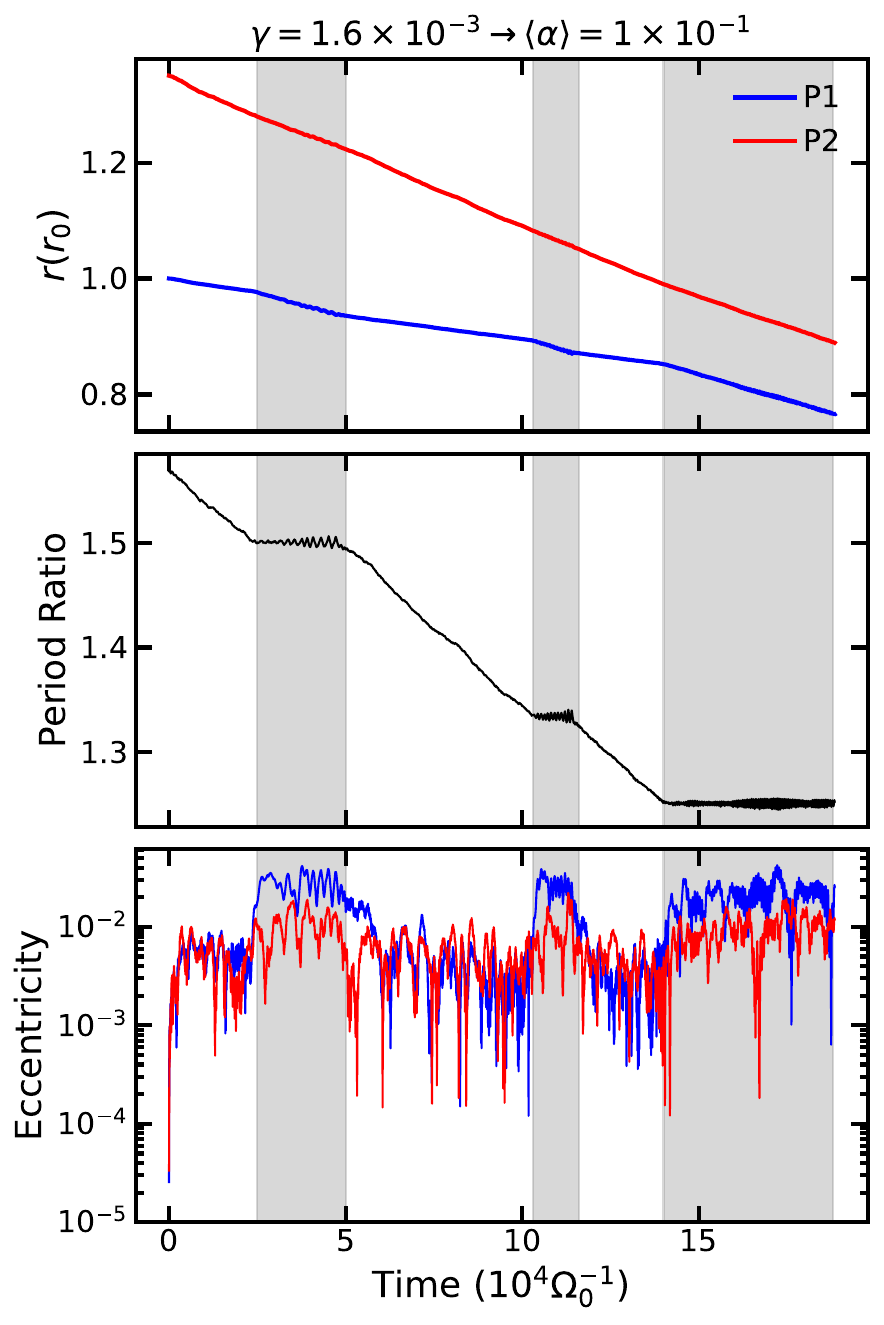}
\caption{Similar to Figure \ref{fig:alpha1e-1}, but for 
strong turbulent case (with $\gamma=1.6\times10^{-3}$, corresponding to an effective viscosity stress parameter $\langle \alpha\rangle=1\times10^{1}$). 
The planet pair traversed a total of three MMR states: 3:2, 4:3, and 5:4 and experienced two instances of resonance breaking.}\label{fig:gamma1e-3}
\end{figure}

\begin{figure*}
\centering
\includegraphics[width=0.95\hsize]{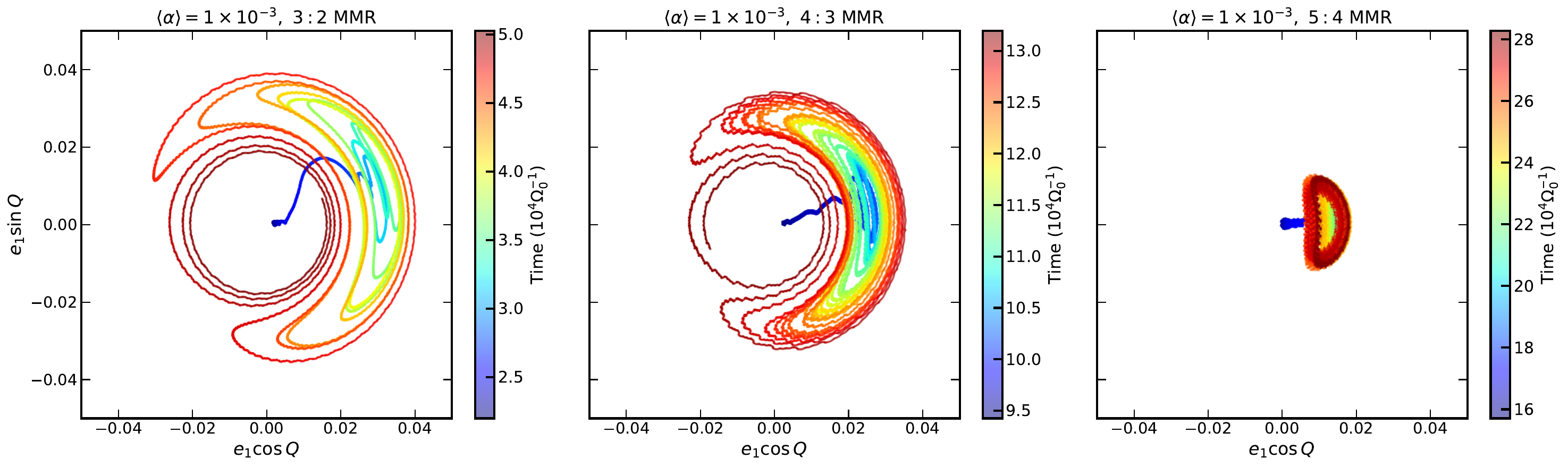}
\caption{Time evolution of the resonance angle phase space of the inner planet for our weak turbulent model $\gamma=1.6\times10^{-4}$. 
The three panels from left to right represent the periods 
when the planet pair become captured into (and escape out of, for unstable capture) 
the 3:2 MMR, 4:3 MMR, and 5:4 MMR, respectively.}\label{fig:gamma1e-4_Qphase}
\end{figure*}

\begin{figure*}
\centering
\includegraphics[width=0.95\hsize]{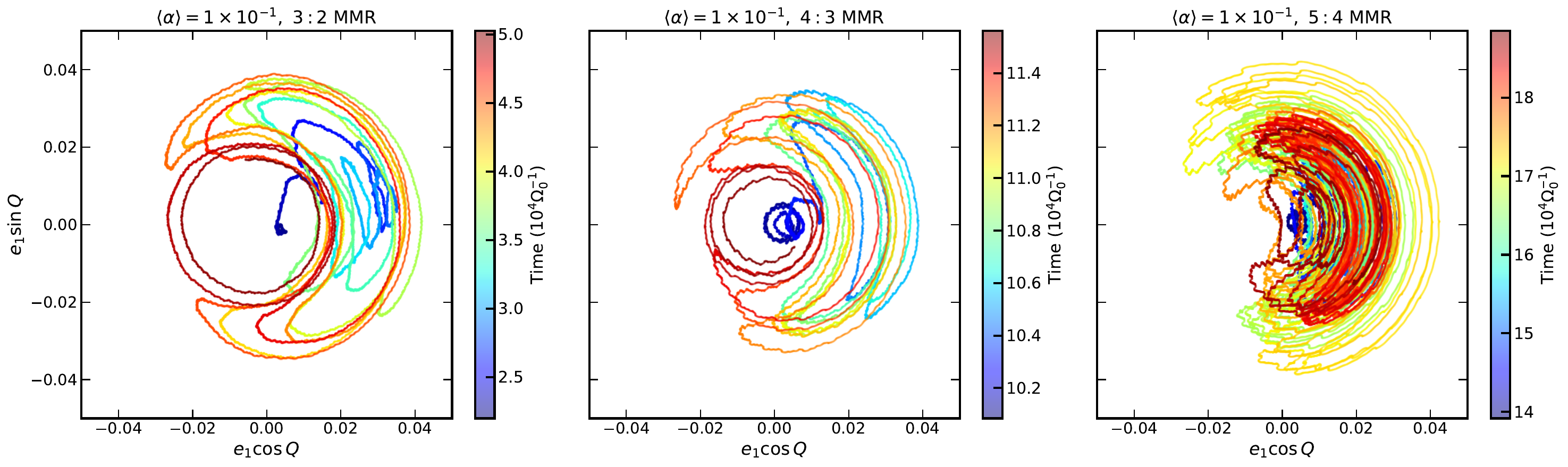}
\caption{Similar to Figure \ref{fig:gamma1e-4_Qphase}, but for strongly turbulent case $\gamma=1.6\times10^{-3}$.}\label{fig:gamma1e-3_Qphase}
\end{figure*}

\subsection{Multi Planets Migration in Turbulent Discs}

We proceed to present result from simulation of multi-planet migration in turbulent environment with the same initial disc mass $\Sigma_0 = 2\times 10^{-5}$. {Scaling $r_0$ to radii within $\lesssim 0.3$ au, 
we expect MRI to be active through sublimation of alkali metals at high temperature \citep{Desch2015}. 
For $r_0 \gtrsim 1$ au and $M_\star = M_\odot$, the column density corresponds to $\Sigma_0 \lesssim 177$ g/cm$^2$, which is mariginaly sufficient
condition for MRI activation via cosmic ray ionization \citep{Gammie1996}.}
As shown in Figure \ref{fig:gamma1e-4} and Figure \ref{fig:gamma1e-3}, 
regardless of whether the turbulence level is weak 
(Figure \ref{fig:gamma1e-4}, $\gamma = 1.6\times10^{-4}$, corresponding to effective $\langle \alpha \rangle=10^{-3}$) 
or strong (Figure \ref{fig:gamma1e-3}, $\gamma = 1.6 \times10^{-3}$, 
corresponding to effective $\langle \alpha \rangle=1\times10^{-1}$), 
the planet pair undergoes 
multiple resonance escape events from 3:2, 4:3 to reach 5:4 at the end of our simulations. 
The behavior of each resonance escape aligns with semi-analytical studies of 
overstability in laminar discs \citep{Goldreich-Schlichting2014}, 
where small perturbations grow to a nonlinear amplitude before they resume their Type I migration and $P_2/P_1$ continues to linearly decrease. 
Similar to Figure \ref{fig:fiducial}, 
we highlight the time interval for each resonance from $t_{\rm in}$ to $t_{\rm out}$ 
(or end of our simulation) to indicate our averaging zones for eccentricities. 
The average values are recorded in Table \ref{tab:resonance}. 

For $\langle \alpha \rangle=10^{-3}$, the equilibrium eccentricity for 3:2 is larger than that for the laminar disc with similar effective $\alpha$ as shown in Figure \ref{fig:gamma1e-4}. 
Furthermore, as evidenced by successive eccentricity plateaus in Figure~\ref{fig:gamma1e-4}, the equilibrium eccentricity progressively decreases with each successive resonance capture (indexed by $j$ for $j+1:j$ MMR) as the planets migrate inward, until the planets are trapped in 5:4 MMR from $t_{\rm in}\approx 1.7\times 10^5\Omega_0^{-1}$ till the end of simulation.
Figure \ref{fig:gamma1e-4_Qphase} shows the time evolution of the resonance angle for this simulation. 
For the 3:2 and 4:3 resonance, 
we observe eccentricity excitation when planet pair enters resonance, then quasi-periodic librations that eventually return to circulation when the pair escapes resonance. 
In contrast, the 5:4 MMR demonstrates significantly more confined motion, 
with the resonance angle eventually restricted to a much smaller region of phase space, 
suggesting a marginally stable resonant capture, 
as we would expect 
from the stronger mutual-interaction due to the closer distance between planets. 

For $\langle \alpha \rangle=10^{-1}$, 
due to the presence of even higher levels of turbulence, 
the growth and decay of eccentricities become more chaotic and it's hard to directly observe how $\langle e\rangle $ changes for each resonance 
from Figure \ref{fig:gamma1e-3} before time-averaging. 
At certain moments, 
the eccentricity of the outer planet even exceeds that of the inner planet.
The evolution of resonance angle is also plotted in the right panel of 
Figure \ref{fig:gamma1e-3_Qphase}. 
However, the 5:4 MMR case shows a significant difference - unlike in the weak turbulence scenario, 
there is no sign of convergence in either eccentricity or resonance angle. 
Instead, the librations observed by the end of the simulation encompass the origin of the phase space, indicating a transition toward circulation.
This implies that if the simulation time were extended, 
the planet pair in this case 
would eventually escape from the 
5:4 MMR to even higher-order resonances.

Qualitatively, 
results from our turbulent simulations imply that active turbulence boosts overstability and helps planets overcome low $j$ MMRs and enter more closely-packed first-order resonances. 
In the next section, 
we will quantitatively show that positive overstability growth rate $s$ 
in these turbulent cases is consistent with the mild increase in 
$\langle e \rangle $ during MMRs.

\section{Analysis}
\label{sec:analysis}

\begin{figure}
\centering
\includegraphics[width=0.85\hsize]{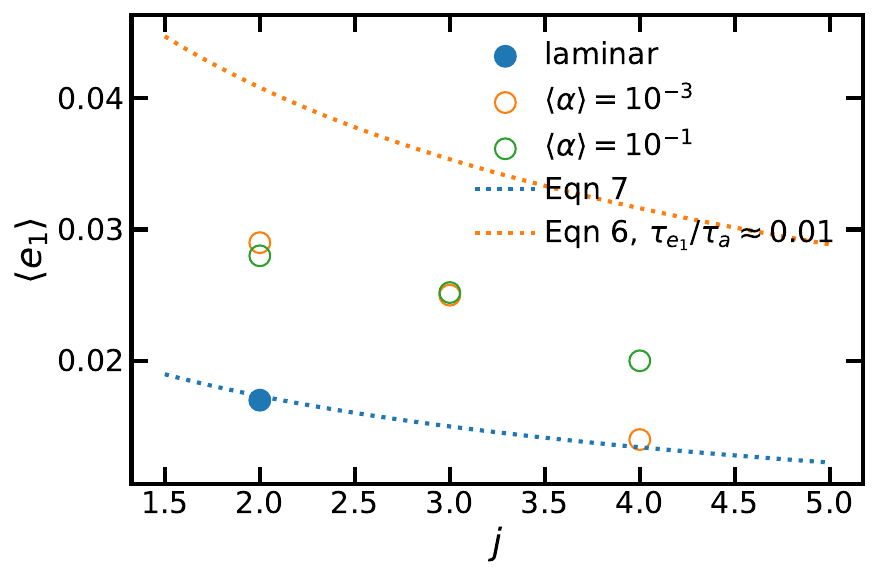}
\caption{Measured average eccentricity of P1 during MMRs for all our simulations, plotted against estimate from the classical laminar context (blue dotted line), as well as combining Equation \ref{eqn:e1full} with $\tau_{e_1}/\tau_a\sim 0.01$ (orange dotted line) as a rough estimate for turbulent cases. }
\label{fig:eeq}
\end{figure}

In Figure \ref{fig:eeq}, we plot P1's average eccentricities during the resonances seen in our simulations 
(one 3:2 resonance for the laminar runs with nearly identical average $e_1$ for different $\alpha$, and three resonances for each turbulent run) 
in hollow circles (values are recorded in Table \ref{tab:resonance}).
In this way, we more clearly see the trend of $e_{1}$ decreasing with $j$ for turbulent cases, 
even though this trend is less significant for $\langle \alpha \rangle=10^{-1}$. 
In laminar discs context, 
for effective semi-major axis and eccentricity damping timescale of 
$\tau_{e}, \tau_{a}$, 
\citet{Goldreich-Schlichting2014, HuangOrmel2023} 
calculated the equilibrium eccentricity of the inner planet in a simplified restricted three-body problem 
(RTBP, assuming the outer planet is fixed and inner planet moves outwards into resonance on timescale of $\tau_{a_1}$) 
to be
\begin{equation}
    e_{\rm eq, RTBP}^2 = {\dfrac{\tau_{e_1}}{2(j + 1)\tau_{a_1}}}.
    \label{eqn:e1}
\end{equation}
An analogous expression has also been obtained for tidally driven orbital migration and 
mean motion resonances of the Galilean satellites of Jupiter \citep{peale1979, linpap1979}.

The generalized expression for more realistic mutually-evolving planet pair migrating \textit{inwards} is given by \citet{Terquem2019}, which can be approximated to order-unity as 
\begin{equation}
    e_{\rm eq}^2 = {\dfrac{\tau_{e_1}}{2(j + 1)\tau_{a}}},
    \label{eqn:e1full}
\end{equation}
where $\tau_a^{-1}: = |\dot{a}_2/a_2| - |\dot{a}_1/a_1|$ is the effective ``approaching" timescale of two planets in the mutually evolving setup. 
Note that $e_{\rm eq}$ can be very close to $e_{\rm eq, RTBP}$ when $\dot{a}_2/a_2 \sim 2 \dot{a}_1/a_1 $ as in our case, 
albeit the physical picture is slightly different. 
Since for the classic type I migration in the subsonic regime in laminar discs we expect $\tau_e/\tau_a \approx h^2$ 
\citep[e.g.][]{Papaloizou2000,Tanaka2002,Li2019,Ida2020} regardless of $j$, 
Equation \ref{eqn:e1} or \ref{eqn:e1full} may be approximated as
\begin{equation}
    e_{\rm eq, laminar}^2 \approx \dfrac{h^2}{j + 1}.
    \label{eqn:eeq,laminar}
\end{equation}

For our laminar disc simulations, 
$j = 2, h = 0.03$ gives $e_{\rm eq, laminar} \approx 0.017$, 
which is very close to our measurement in both $\alpha= 10^{-3}$ and $\alpha= 10^{-1}$ 
simulations
even if we do not explicitly invoke measurements of damping timescale. We plot $e_{\rm eq, laminar}$ across $j$ with a blue dotted line in Figure \ref{fig:eeq}.
For our turbulent cases, 
however, 
the eccentricities are systematically larger  with stronger instability, 
and does 
not follow a $\propto (j+1)^{-1/2}$ decay as we might expect from a laminar disc.

To assess potential offsets in numerical measurements from expectations based on Equation \ref{eqn:e1full}, 
we need to 
provide some estimates of $\tau_e$ and $\tau_a$ for disc dissipation for our turbulent simulations. 
While it's easy to average over the periods when planet pairs are undergoing Type I migration out of resonance to obtain $\tau_a \approx 4\times 10^5 \Omega_0^{-1}$, 
measurement of $\tau_e$ is tricky since $\dot{e}/e$ 
can fluctuate very strongly when $e$ becomes negligible during Type I migration. 
The most reliable approach measuring $\tau_e$ is to average over a relatively short period $(\sim 10^4 \Omega_{0}^{-1})$ 
right after $t_{\rm out}$, 
when eccentricity follows a nearly exponential decay and $\dot{e}/e$ is very well-defined \citep{Hands-Alexander2018}. Using this method, 
we measure $\tau_{e_1} \approx 4\times 10^3 \Omega_0^{-1}$ for both runs, 
yielding an estimate of $\tau_{e_1}/\tau_a \sim 0.01$.

Finally, in Figure \ref{fig:eeq}, we plot the theoretical equilibrium eccentricity 
by combining the measured value of $\tau_{e_1}/\tau_a \sim 0.01$ for turbulent discs
with Equation \ref{eqn:e1full} (orange dotted line). 
We see this could serve as an upper limit for theoretical estimate, 
though it may not be an accurate approximation. The overestimation compared to simulation result
suggests that other non-linear effects may be providing feedback
limiting the eccentricity, which implies that 
Equation \ref{eqn:e1full} may not be able to capture all the dynamics of eccentricity excitation 
when active turbulence is present.

\begin{figure}
\centering
\includegraphics[width=0.85\hsize]{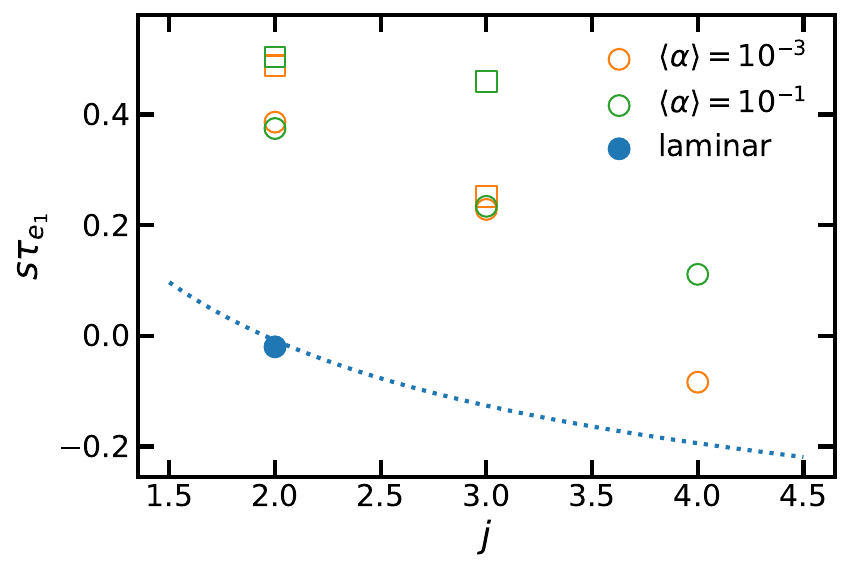}
\caption{The dimensionless growth rate calculated from average eccentricity of P1 during MMRs for all our simulations, 
plotted against estimate from the classical laminar context (blue dotted line) which predicts laminar case to be stable. We also plot direct measurements for certain MMRs in square symbols. }
\label{fig:seq}
\end{figure}

\begin{figure*}
\centering
\includegraphics[width=0.47\hsize]{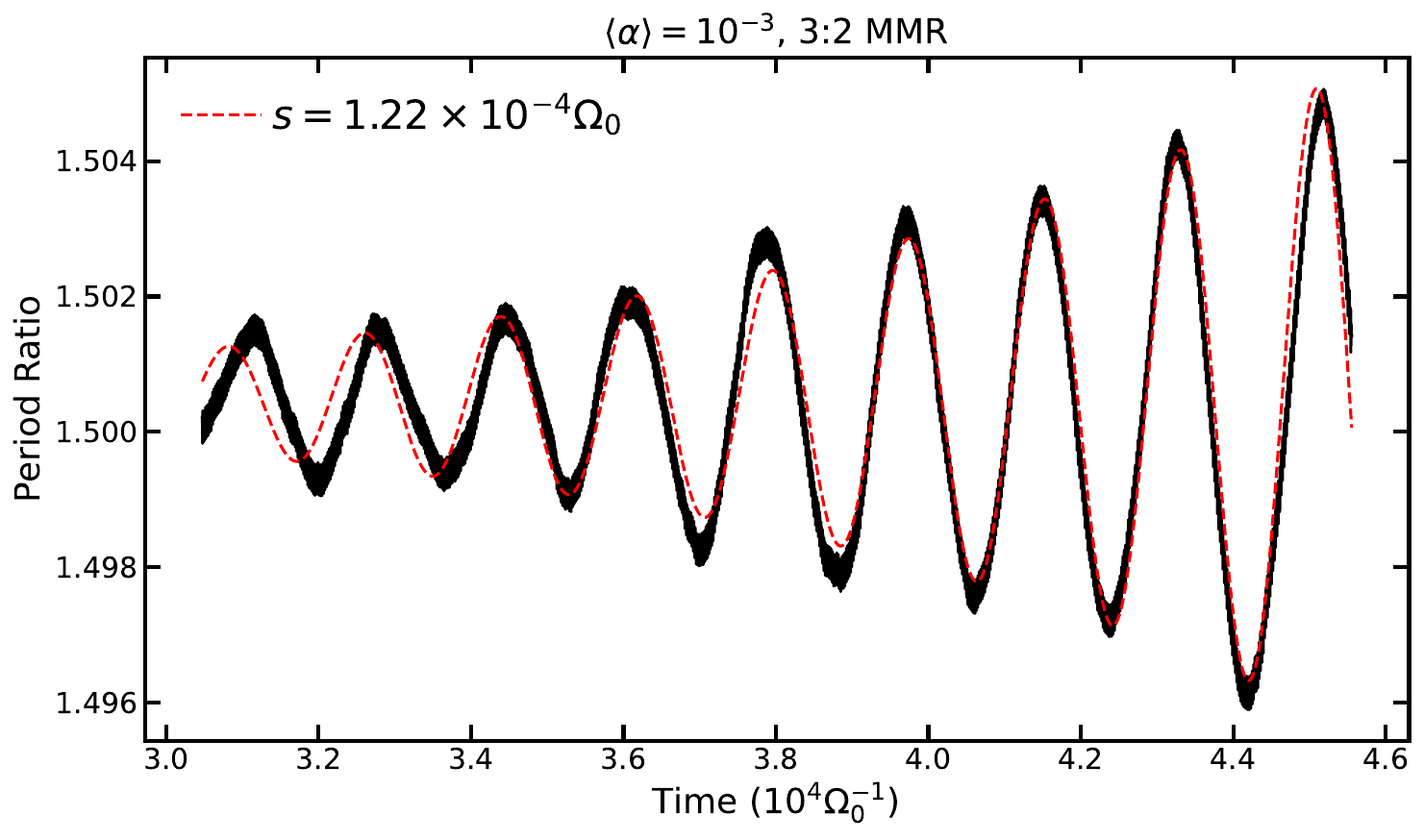}
\includegraphics[width=0.47\hsize]{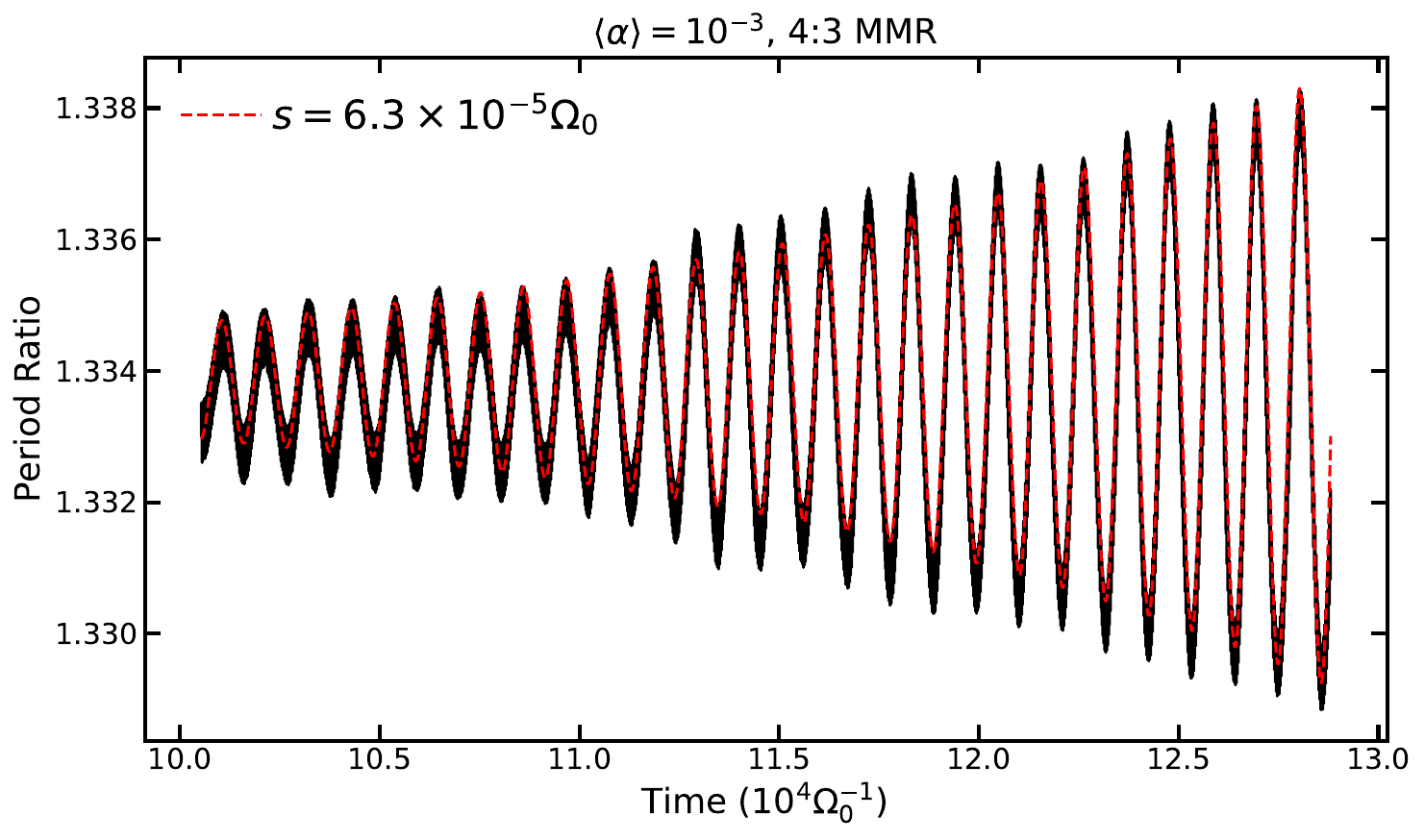}
\includegraphics[width=0.47\hsize]{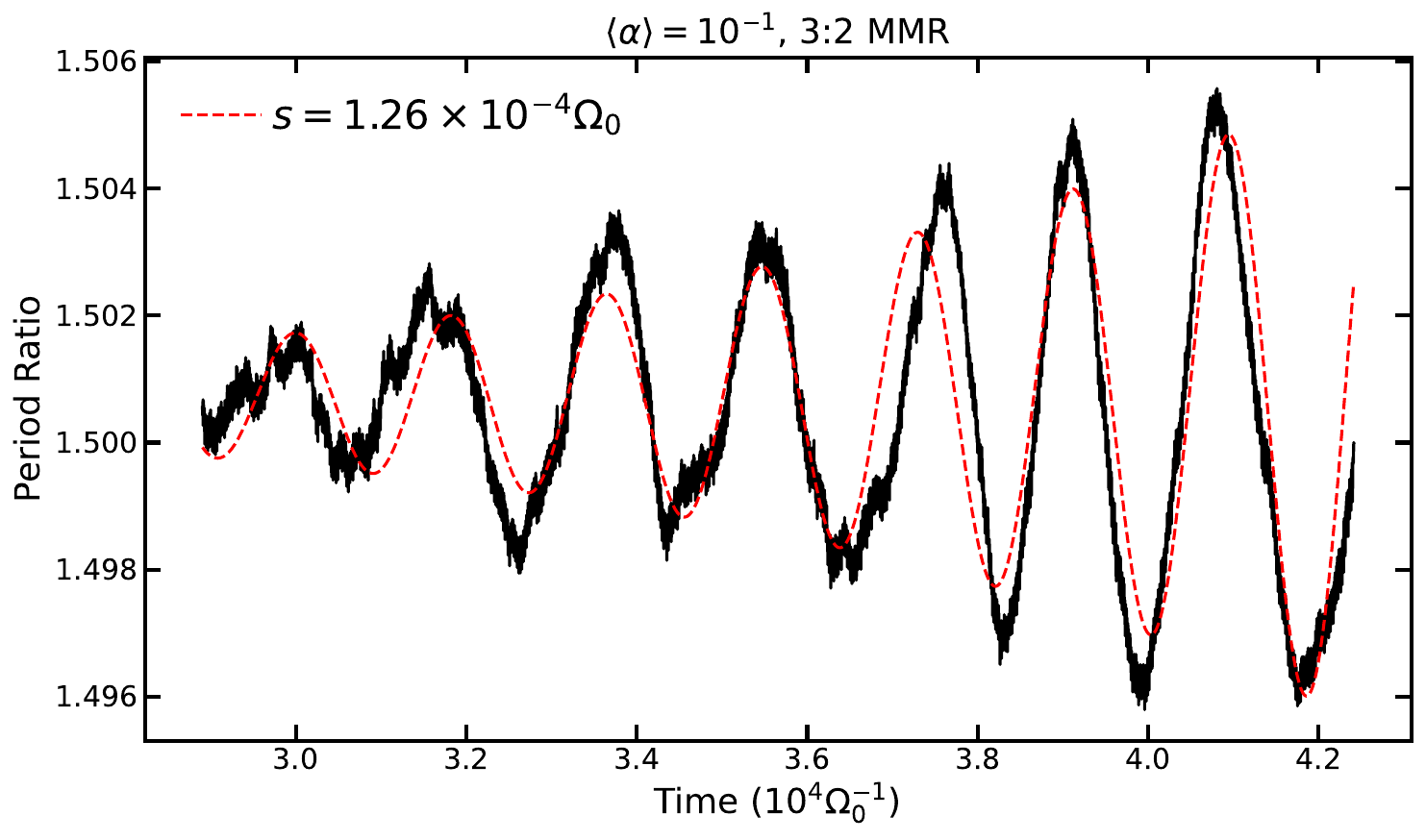}
\includegraphics[width=0.47\hsize]{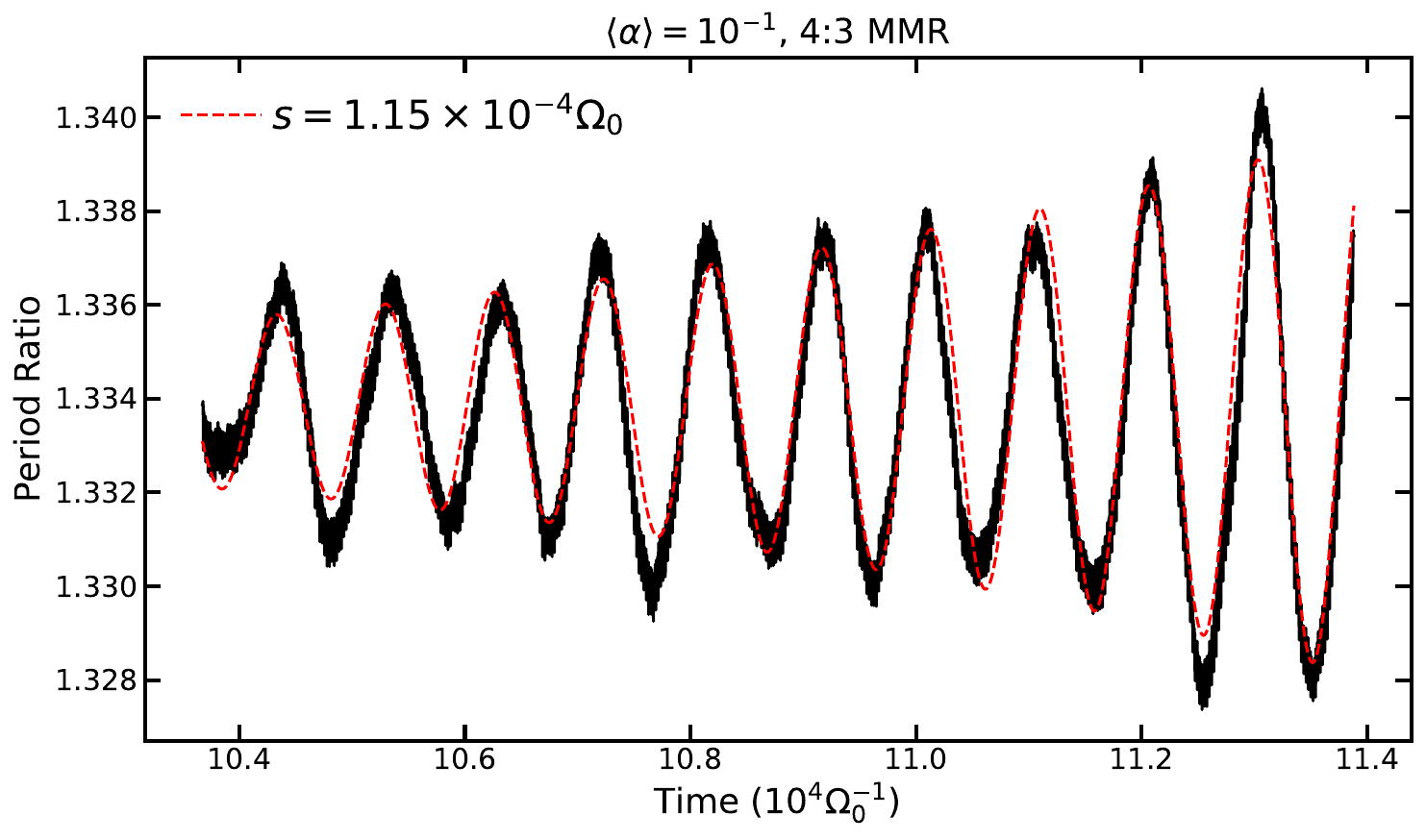}
\caption{Best fits to the libration during MMRs with growth rates labeled for two turbulence runs.}
\label{fig:sfit}
\end{figure*}

In the RTBP approximation \citep{Goldreich-Schlichting2014}, 
linear perturbations in orbital parameters (e.g. eccentricity, 
period ratio) evolve with time as $\delta \propto \exp (s + i\omega) t$ when planets are in resonance, where 
\begin{equation}
    \omega/\Omega = \sqrt{3 j^2 f_j \tilde{\alpha} q_2 e_{\rm eq}  + \left(\dfrac{f_j \tilde{\alpha} q_2}{e_{\rm eq}}\right)^2},
    \label{eqn:omega}
\end{equation}
and 
\begin{equation}
    s \tau_e = \left(\frac{\Omega}{\omega}\right)^2\left(3 j f_j \tilde{\alpha} q_2 e_{\mathrm{eq}}-\left(\frac{{ f_j \tilde{\alpha} q_2}}{e_{\mathrm{eq}}}\right)^2\right).
    \label{eqn:staue}
\end{equation}
Here $\tilde{\alpha}= (j/(j + 1))^{2/3}$ is the semi-major axis ratio, 
and $f_j$ is an order-unity coefficients dependent on $j$ \footnote{In fact, $\tilde{\alpha} f_j \approx 0.8 j$ which is utilized by \citet{Goldreich-Schlichting2014}.}. One can see clearly that $s \tau_e $ is a sharply decreasing function of $e_{\mathrm{eq}}$.

We plot the predicted $s\tau_{e_1}$ in Figure \ref{fig:seq} by substituting measured $\langle e_1\rangle $ into Equation \ref{eqn:staue} (corresponding to circles of similar color in Figure \ref{fig:eeq}). 
To compare, 
we also plot the $s\tau_{e_1}$ profile calculated from Equation \ref{eqn:eeq,laminar} 
expected for laminar discs, shown as the dotted blue line.
We find that $s<0$ for $j\geq2$, 
which suggests there will be no overstability as resonances become more and more stable for higher $j$. 
For turbulent cases, 
we see a systematic increase in $s$ solely due to larger $e_{\rm eq}$ values. This difference is sufficient to shift $s$ to positive values at $j = 2, 3$ 
for $\langle \alpha \rangle=10^{-3}$ and at all $j = 2, 3, 4$ 
for $\langle \alpha \rangle=10^{-1}$. 

Qualitatively, 
the sign change of expected $s\tau_{e_1}$ alone explains 
why turbulent cases are more prone to overstability, 
regardless of the value of $\tau_{e_1}$.
To quantitatively compare with numerical results, 
we extract period ratio data from our turbulent runs during libration in MMRs 
and explicitly determine the perturbation growth rate. 
The exponential growth is apparent at least for the 3:2 and 4:3 resonances, 
where libration amplitude increases nearly monotonically (see black solid lines in Figure \ref{fig:sfit}). 
However, 
it is not well-defined for the 5:4 MMR, where oscillation amplitude 
rises and falls chaotically towards the end of both simulations.

Simple Fourier transform for time evolution of period ratio during the 3:2 and 4:3 resonances will inform us 
that the peak libration frequency $\omega \sim 0.003\Omega_0$ for all cases, 
aligning with estimation from Equation \ref{eqn:omega} which is expected to be quite insensitive to $e_{\rm eq}$. 
However, 
measurement of $s$ this way is quite uncertain since the entire time series from capture to escape only 
spans roughly one e-folding timescale 
for the amplitude's exponential growth \citep{Goldreich-Schlichting2014}, 
and 
Fourier spectrum becomes unreliable at low frequencies 
when the period approaches 
the total duration of the time series. We find it more straightforward to apply a nonlinear least-squares fit to the data, optimizing 
$\omega$ 
(which turns out to be consistent with results from Fourier transform) and $s$ 
jointly with the libration's 
phase and amplitude. We plot best fits against data for 3:2 and 4:3 MMRs in our turbulent cases in Figure \ref{fig:sfit}. 
We further multiply the values of best-fit $s$ by $\tau_{e_1} \sim 4\times 10^3 \Omega_0^{-1}$ and plot the 4 ``measured" values in Figure \ref{fig:seq} with square symbols. By comparison, 
Equation \ref{eqn:staue} (circles) proves to be a sufficient approximation for $\langle \alpha \rangle=10^{-3}$ as long as we substitute in consistent measurements of equilibrium eccentricity. 
For $\langle \alpha \rangle=10^{-1}$, 
the values differs at 4:3 resonance by a factor of 2
but even so the measured $s$ still exhibits a  decreasing trend with higher $j$. This suggests that, if migration continued indefinitely, 
the planets would ideally become stabilized in a much tighter resonance. 

Finally, we note that no such measurements can be made for laminar cases as significant librations are absent. 
Nevertheless,  
the simple fact 
that the 3:2 MMR become stable 
is indeed 
consistent with $s<0$ for $j \geq 2$ along the blue dotted line in Figure \ref{fig:seq}. 

\begin{table*}
\centering
\begin{tabular}{l|c|c|c|c|c|c|c|c|}
\hline
\hline
 Run &$ \alpha  =10^{-3}$ & $ \alpha  =10^{-1}$ &  \multicolumn{3}{c|}{$\gamma=1.6\times 10^{-4}\rightarrow\langle \alpha \rangle =10^{-3}$} & \multicolumn{3}{c|}{$\gamma=1.6\times 10^{-3}\rightarrow\langle \alpha \rangle =10^{-1}$} \\
\hline
resonance & 3:2 & 3:2  & 3:2 & 4:3 & 5:4 & 3:2 & 4:3 & 5:4 \\
\hline
$\langle e_1 \rangle$ & 0.018 & 0.018 & 0.029 & 0.025 & 0.014 & 0.028 & 0.025 & 0.020\\
\hline
$\langle e_2 \rangle$ & 0.006 & 0.006 & 0.008 & 0.007 & 0.004& 0.010 & 0.009 & 0.009 \\
\hline
\end{tabular}
\caption{Measured equilibrium eccentricity during MMRs for all our simulations.}
\label{tab:resonance}
\end{table*}

\section{Summary and Implications}
\label{sec:summary}

In this paper, 
we apply hydrodynamical simulations of actively migrating planet pairs with 
laminar viscosity prescription versus active turbulence, 
to investigate the stability of first-order MMRs. 
In simulations with laminar viscosity $\alpha = 10^{-3} - 10^{-1}$, 
low-mass planet pairs become trapped into a perfect 3:2 resonance. 
With more realistic modeling of active turbulence, 
the 3:2 resonance becomes overstable. This overstability 
causes the period ratio and eccentricity to librate over time with a positive growth rate $s$ in amplitude, 
eventually resulting in resonance escape. 
The existence of these unstable modes $s>0$ are due to higher equilibrium eccentricities in a turbulent disc, 
albeit $s$ 
decreases as planets migrate into more stable, closer-in resonances, 
a trend that we also expect in classical theory for laminar discs.
For effective $\langle \alpha \rangle = 10^{-3}$, the planet pairs migrate past the 3:2 and 4:3 MMR, eventually stabilizing at the 5:4 MMR. 
For effective $\langle \alpha \rangle = 10^{-1}$, even the 5:4 MMR would become unstable by the end of our simulation, and we expect the planet pair to eventually escape given sufficient time.

The direct conclusion from our analysis indicates that realistic turbulence expands the parameter space for overstability, 
causing planet pairs to migrate into tighter resonances compared to laminar discs. 
While we present the first hydrodynamic simulation for this comparison, 
a similar trend is observed in \citet[][see their Figure 9]{Izidoro2017}, where effect of turbulence is modeled through a turbulent-potential prescription in their N-body simulations. 
However, such incorporation of turbulence in N-body simulations \citep{Ogihara2007,Izidoro2017,Secunda2019} involves directly applying potential terms calibrated from simulations \citep{Laughlin2004,BaruteauLin2010} (namely Equation \ref{eqn:phiturb}) \textit{on the planets}, 
which can be physically different from what is initially envisioned from the simulations: 
these potential terms are supposed to be applied to the disc gas 
to generate eddies in the surface density profile, 
mimicking the effects of effective Reynolds stress 
and the associated turbulent power spectrum. These eddies, in turn, influences the migration/damping torque exerted on the planets through modification of location and strength of Lindblad resonances. As we suggested in \citet{Wu2024chaotic}, 
a more realistic and straightforward way may be to apply stochastic torques  \citep{Rein-Papaloizou2009,Paardekooper2013,Hands2014}, 
with dispersion as functions of disc parameters calibrated by turbulent simulations. 
Notably, 
a conclusion in \citet{Izidoro2017} is that the 
dichotomy between period ratio distribution in turbulent and laminar disc at the end of disc phase is small enough, 
such that it will be smeared out after reshaping of planet architecture by dynamical instability. 
Whether this relies on 
their specific treatment of turbulence remains to be investigated by updated population studies. 

In terms of the deviation from commensurability, 
\begin{equation}
    \Delta : = \dfrac{P_2/P_1}{(j+1)/j}-1,
    \label{eqn:delta}
\end{equation}
libration in our simulation typically results in $\Delta \sim 0.5\%$, 
which, although much larger than measured value for MMR in the laminar disc ($\Delta \sim 0.05\%$), 
is not enough to explain certain observed systems with $\Delta \sim 1-3\%$ \citep{Dai2024} by itself. 
This value may serve as an adequate initial seed for resonant repulsion that eventually expands to values consistent with observation through post-PPD evolution \citep{Xu-Dai2025} or other mechanisms \citep[e.g.,][]{PapaloizouTerquem2010, LithwickWu2012,Batygin2013,Delisle2014,Choksi2020}. 
{The prevalence of near-resonant planets in TESS systems close to 3:2 and 2:1 MMRs  \citep{Dai2024} 
may suggest low turbulence in their birth environments, 
with certain sources' natal discs being quite laminar ($\Delta \sim 10^{-4}$), 
such as TOI-1136 \citep{Dai2023}, Kepler 60 and Kepler 223 \citep{Fabrycky2014}. 
Reversely, 
relatively lower occurence rate of planets near tighter (4:3 or 5:4) resonances does not necessarily indicate the scarcity of high-turbulence discs. 
Although these tighter resonances are more stable during migration, 
they are also more susceptible to resonance crossing 
and dynamical instability after disc dispersal \citep{Lammers2024}, 
which can disrupt MMRs. 
Nevertheless, some exceptions are emphasized by
\citet{RixinLi2024}, 
who show that merger events due to dynamical instability might not 
eliminate near-resonant planet pairs but instead contribute to a moderate widening of $\Delta$ for surviving planets. }

In a steady state, the accretion rate through the disc ${\dot M}_d = 3 \pi \Sigma \nu$
is independent of $r$. Normalized by the values of $h_0 (=0.03)$ and 
$\Sigma_0 (=2 \times 10^{-5})$ we have adopted in the fiducial model, 
\begin{equation}
    {\dot M}_d = 2.7 \times 10^{-8}
    \alpha \left( {h_0 \over 0.03} \right)^2 \left( {\Sigma_0 \over 2 \times 10^{-5}} \right) 
    \left( {M_\star \over M_\odot} {1~{\rm au} \over 
    r_0} \right)^{3/2} {M_\odot \over {\rm yr}}.
\end{equation}
Even with a relatively large $\alpha (\sim 0.1)$, 
the parameters used in the fiducial model are appropriate for the late stages of disc evolution 
around weak-line T Tauri stars.  For the range of observed accretion rate onto 
classical T Tauri stars \citep{hartmann1998}, larger values of $\Sigma_0$ are relevant.
Such models may to lead more compact mean motion resonances (see the laminar model
in Appendix A) between planets more massive than those simulated in the fiducial model.  
In a follow-up investigation,
we will consider the possibility that, in turbulent discs around classical T Tauri stars, 
mean motion resonances between super-Earths may overlap with each 
other and the potential implications of resulting dynamical instability.

We also note that our treatment of active turbulence still differs from realistic MRI and GI generated 
from simulations incorporating magnetic fields and self-gravity \citep{Simon2012, deng2020}. 
Analogous investigations of MMR capture and detachment 
in such environments are warranted and may not necessarily yield the same quantitative results.
Nevertheless, our framework of comparing equilibrium eccentricity and 
growth rates of overstability during resonances
with benchmark for analytical theory/laminar discs remains valuable. 

{Finally, we believe that with the upcoming operation of longer-baseline radio facilities in the coming decades, such as the Square Kilometre Array (SKA) and next generation Very Large Array (ngVLA), observational constraints on turbulence within PPDs and the ability to image substructures will continue to improve. These observations will help further constrain the turbulence strength in PPDs, particularly in the inner disc regions \citep{Ricci-etal.2018,Wu-etal.2024-SKA}, and may even capture unusual substructures caused by planets migration or MMRs \citep{Wu_2023}.}


\section*{Acknowledgements}
We thank Pablo Ben{\'\i}tez-Llambay, Hongping Deng, Shuo Huang, Rixin Li and Linghong Lin for helpful discussions and the anonymous referee for a thorough reading and positive report. Y.W. acknowledges the EACOA Fellowship awarded by the East Asia Core Observatories Association. Y.P.L. is supported in part by the Natural Science Foundation of China (grants 12373070, and 12192223), the Natural Science Foundation of Shanghai (grant NO. 23ZR1473700). R.A. acknowledges funding from the Science \& Technology Facilities Council (STFC) through Consolidated Grant ST/W000857/1.
This research used the DiRAC Data Intensive service at Leicester, operated by the University of Leicester IT Services, which forms part of the STFC DiRAC HPC Facility (\href{www.dirac.ac.uk}{www.dirac.ac.uk}). The calculations have made use of the High Performance Computing Resource in the Core Facility for Advanced Research Computing at Shanghai Astronomical Observatory.

\section*{Data availability}

The data obtained in our simulations can be made available on reasonable request to the corresponding author. 

\appendix

\section{Bypassing resonances in a heavier disc}\label{appendix: disc-model}

\begin{figure}
\centering
\includegraphics[width=1\hsize]{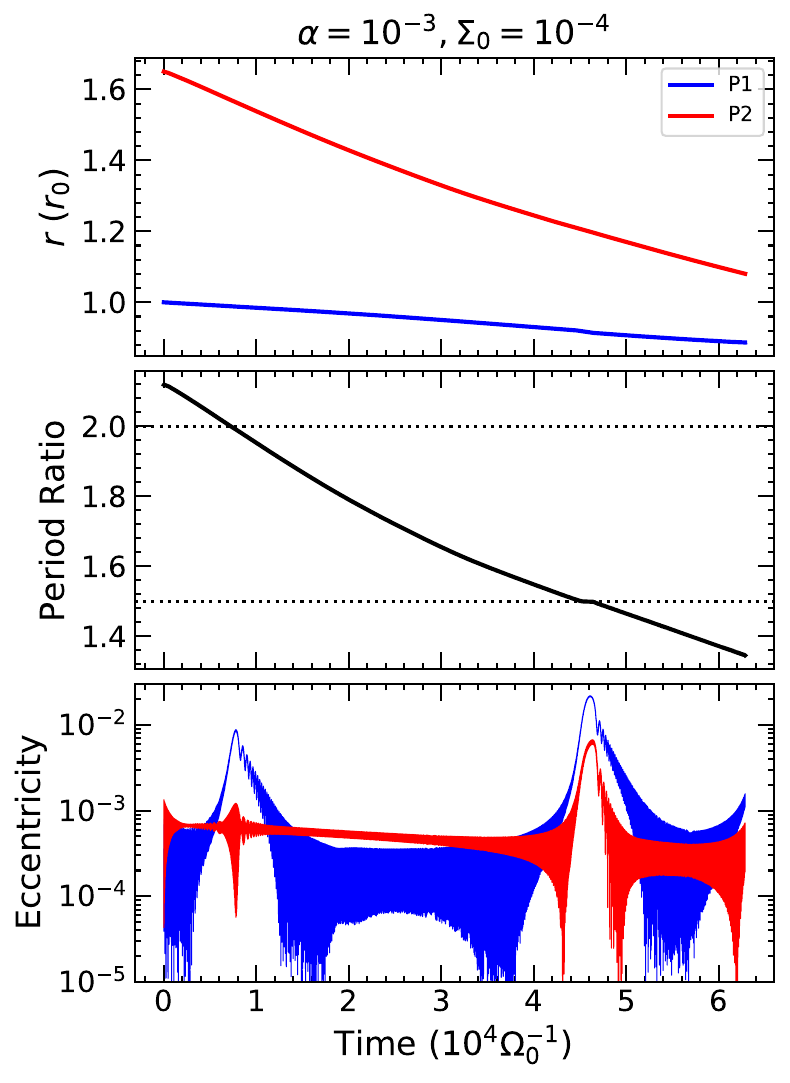}
\caption{Orbital parameter evolution for additional laminar simulation with $\alpha = 10^{-3}$ and $\Sigma_0  = 10^{-4}$. We demonstrate that planet pairs can migrate past the 2:1  and 3:2 MMR 
(dotted lines) 
without being captured when the migration speed is high due to a heavier disc. }\label{fig:Sigma1e-4}
\end{figure}

In Figure \ref{fig:Sigma1e-4}, 
we present time evolution of orbital parameters for a 
laminar viscous simulation with $\alpha=10^{-3}$ and
$\Sigma_0 = 10^{-4} [M_\star/r_0^2]$. 
We start with P2 at a semi-major axis of $ 1.65 r_0$, 
and the planet pair migrated past both the 2:1 and 3:2 resonances without capture. 
For we can measure that $\tau_a \sim 10^5 \Omega_0^{-1}$, 
shorter than the critical damping timescale required for slow-migration and capture into resonance \citep{Lin2025}:

\begin{equation}
    \tau_{a, \rm crit} \approx \dfrac{1}{2(3j)^{1/3}}\dfrac{1}{(f_j \tilde{\alpha} q_2)^{4/3}}\Omega_0^{-1},
\end{equation}
which equals to $ 3.97\times 10^{5}\Omega_0^{-1} $ for 3:2 MMR and $ 1.26\times 10^{6}\Omega_0^{-1} $ for the much weaker 2:1 MMR.
Therefore, 
resonance capture is unlikely in this laminar scenario due to rapid migration. 
Theoretically it could occur for $5:4$ MMR since $\tau_{a, \rm crit}$ decreases with $j$, but we do not extend our simulation longer to confirm as this is not the focus of our study. 
The value of $\tau_{a, \rm crit}$ at 3:2 MMR is consistent with our fiducial case, 
where the surface density is five times lower, resulting in a migration timescale approximately five times longer and 
satisfies the slow migration criterion.

\bibliographystyle{mnras}
\bibliography{multi-planet-turb} 

\begin{thebibliography}{}
\makeatletter
\relax
\def\mn@urlcharsother{\let\do\@makeother \do\$\do\&\do\#\do\^\do\_\do\%\do\~}
\def\mn@doi{\begingroup\mn@urlcharsother \@ifnextchar [ {\mn@doi@}
  {\mn@doi@[]}}
\def\mn@doi@[#1]#2{\def\@tempa{#1}\ifx\@tempa\@empty \href
  {http://dx.doi.org/#2} {doi:#2}\else \href {http://dx.doi.org/#2} {#1}\fi
  \endgroup}
\def\mn@eprint#1#2{\mn@eprint@#1:#2::\@nil}
\def\mn@eprint@arXiv#1{\href {http://arxiv.org/abs/#1} {{\tt arXiv:#1}}}
\def\mn@eprint@dblp#1{\href {http://dblp.uni-trier.de/rec/bibtex/#1.xml}
  {dblp:#1}}
\def\mn@eprint@#1:#2:#3:#4\@nil{\def\@tempa {#1}\def\@tempb {#2}\def\@tempc
  {#3}\ifx \@tempc \@empty \let \@tempc \@tempb \let \@tempb \@tempa \fi \ifx
  \@tempb \@empty \def\@tempb {arXiv}\fi \@ifundefined
  {mn@eprint@\@tempb}{\@tempb:\@tempc}{\expandafter \expandafter \csname
  mn@eprint@\@tempb\endcsname \expandafter{\@tempc}}}

\bibitem[\protect\citeauthoryear{{Afkanpour}, {Ataiee}, {Ziampras}, {Penzlin},
  {Sfair}, {Sch{\"a}fer}, {Kley}  \& {Schlichting}}{{Afkanpour}
  et~al.}{2024}]{Afkanpour2024MMR}
{Afkanpour} Z.,  {Ataiee} S.,  {Ziampras} A.,  {Penzlin} A. B.~T.,  {Sfair} R.,
   {Sch{\"a}fer} C.,  {Kley} W.,   {Schlichting} H.,  2024, \mn@doi [\aap]
  {10.1051/0004-6361/202348826}, \href
  {https://ui.adsabs.harvard.edu/abs/2024A&A...686A.277A} {686, A277}

\bibitem[\protect\citeauthoryear{{Alexander}, {Rosotti}, {Armitage}, {Herczeg},
  {Manara}  \& {Tabone}}{{Alexander} et~al.}{2023}]{Alexander-etal.2023}
{Alexander} R.,  {Rosotti} G.,  {Armitage} P.~J.,  {Herczeg} G.~J.,  {Manara}
  C.~F.,   {Tabone} B.,  2023, \mn@doi [\mnras] {10.1093/mnras/stad1983}, \href
  {https://ui.adsabs.harvard.edu/abs/2023MNRAS.524.3948A} {524, 3948}

\bibitem[\protect\citeauthoryear{{Ataiee} \& {Kley}}{{Ataiee} \&
  {Kley}}{2021}]{Ataiee-Kley2021}
{Ataiee} S.,  {Kley} W.,  2021, \mn@doi [\aap] {10.1051/0004-6361/202038772},
  \href {https://ui.adsabs.harvard.edu/abs/2021A&A...648A..69A} {648, A69}

\bibitem[\protect\citeauthoryear{{Baruteau} \& {Lin}}{{Baruteau} \&
  {Lin}}{2010}]{BaruteauLin2010}
{Baruteau} C.,  {Lin} D.~N.~C.,  2010, \mn@doi [\apj]
  {10.1088/0004-637X/709/2/759}, \href
  {https://ui.adsabs.harvard.edu/abs/2010ApJ...709..759B} {709, 759}

\bibitem[\protect\citeauthoryear{{Batygin} \& {Morbidelli}}{{Batygin} \&
  {Morbidelli}}{2013}]{Batygin2013}
{Batygin} K.,  {Morbidelli} A.,  2013, \mn@doi [\aj]
  {10.1088/0004-6256/145/1/1}, \href
  {https://ui.adsabs.harvard.edu/abs/2013AJ....145....1B} {145, 1}

\bibitem[\protect\citeauthoryear{{Batygin} \& {Petit}}{{Batygin} \&
  {Petit}}{2023}]{Batygin2023}
{Batygin} K.,  {Petit} A.~C.,  2023, \mn@doi [\apjl]
  {10.3847/2041-8213/acc015}, \href
  {https://ui.adsabs.harvard.edu/abs/2023ApJ...946L..11B} {946, L11}

\bibitem[\protect\citeauthoryear{{Beckwith}, {Armitage}  \& {Simon}}{{Beckwith}
  et~al.}{2011}]{Beckwith2011}
{Beckwith} K.,  {Armitage} P.~J.,   {Simon} J.~B.,  2011, \mn@doi [\mnras]
  {10.1111/j.1365-2966.2011.19043.x}, \href
  {https://ui.adsabs.harvard.edu/abs/2011MNRAS.416..361B} {416, 361}

\bibitem[\protect\citeauthoryear{{Ben{\'\i}tez-Llambay} \&
  {Masset}}{{Ben{\'\i}tez-Llambay} \& {Masset}}{2016}]{FARGO3D}
{Ben{\'\i}tez-Llambay} P.,  {Masset} F.~S.,  2016, \mn@doi [\apjs]
  {10.3847/0067-0049/223/1/11}, \href
  {https://ui.adsabs.harvard.edu/abs/2016ApJS..223...11B} {223, 11}

\bibitem[\protect\citeauthoryear{{Benz}, {Ida}, {Alibert}, {Lin}  \&
  {Mordasini}}{{Benz} et~al.}{2014}]{benz2014}
{Benz} W.,  {Ida} S.,  {Alibert} Y.,  {Lin} D.,   {Mordasini} C.,  2014, in
  {Beuther} H.,  {Klessen} R.~S.,  {Dullemond} C.~P.,   {Henning} T.,  eds,
  Protostars and Planets VI. pp 691--713 (\mn@eprint {arXiv} {1402.7086}),
  \mn@doi{10.2458/azu_uapress_9780816531240-ch030}

\bibitem[\protect\citeauthoryear{{Carr}, {Tokunaga}  \& {Najita}}{{Carr}
  et~al.}{2004}]{Carr-etal.2004}
{Carr} J.~S.,  {Tokunaga} A.~T.,   {Najita} J.,  2004, \mn@doi [\apj]
  {10.1086/381356}, \href
  {https://ui.adsabs.harvard.edu/abs/2004ApJ...603..213C} {603, 213}

\bibitem[\protect\citeauthoryear{{Chen} \& {Lin}}{{Chen} \&
  {Lin}}{2023}]{ChenLin2023}
{Chen} Y.-X.,  {Lin} D. N.~C.,  2023, \mn@doi [\mnras] {10.1093/mnras/stad992},
  \href {https://ui.adsabs.harvard.edu/abs/2023MNRAS.522..319C} {522, 319}

\bibitem[\protect\citeauthoryear{{Chen}, {Li}, {Li}  \& {Lin}}{{Chen}
  et~al.}{2020}]{Chen2020}
{Chen} Y.-X.,  {Li} Y.-P.,  {Li} H.,   {Lin} D. N.~C.,  2020, \mn@doi [\apj]
  {10.3847/1538-4357/ab9604}, \href
  {https://ui.adsabs.harvard.edu/abs/2020ApJ...896..135C} {896, 135}

\bibitem[\protect\citeauthoryear{{Chiang} \& {Laughlin}}{{Chiang} \&
  {Laughlin}}{2013}]{Chiang2013}
{Chiang} E.,  {Laughlin} G.,  2013, \mn@doi [\mnras] {10.1093/mnras/stt424},
  \href {https://ui.adsabs.harvard.edu/abs/2013MNRAS.431.3444C} {431, 3444}

\bibitem[\protect\citeauthoryear{{Chiang} \& {Youdin}}{{Chiang} \&
  {Youdin}}{2010}]{Chiang2010}
{Chiang} E.,  {Youdin} A.~N.,  2010, \mn@doi [Annual Review of Earth and
  Planetary Sciences] {10.1146/annurev-earth-040809-152513}, \href
  {https://ui.adsabs.harvard.edu/abs/2010AREPS..38..493C} {38, 493}

\bibitem[\protect\citeauthoryear{{Choksi} \& {Chiang}}{{Choksi} \&
  {Chiang}}{2020}]{Choksi2020}
{Choksi} N.,  {Chiang} E.,  2020, \mn@doi [\mnras] {10.1093/mnras/staa1421},
  \href {https://ui.adsabs.harvard.edu/abs/2020MNRAS.495.4192C} {495, 4192}

\bibitem[\protect\citeauthoryear{{Cleaver}, {Hartmann}  \& {Bae}}{{Cleaver}
  et~al.}{2023}]{Cleaver-etal.2023}
{Cleaver} J.,  {Hartmann} L.,   {Bae} J.,  2023, \mn@doi [\mnras]
  {10.1093/mnras/stad1784}, \href
  {https://ui.adsabs.harvard.edu/abs/2023MNRAS.523.5522C} {523, 5522}

\bibitem[\protect\citeauthoryear{{Dai} et~al.,}{{Dai} et~al.}{2023}]{Dai2023}
{Dai} F.,  et~al., 2023, \mn@doi [\aj] {10.3847/1538-3881/aca327}, \href
  {https://ui.adsabs.harvard.edu/abs/2023AJ....165...33D} {165, 33}

\bibitem[\protect\citeauthoryear{{Dai} et~al.,}{{Dai} et~al.}{2024}]{Dai2024}
{Dai} F.,  et~al., 2024, \mn@doi [\aj] {10.3847/1538-3881/ad83a6}, \href
  {https://ui.adsabs.harvard.edu/abs/2024AJ....168..239D} {168, 239}

\bibitem[\protect\citeauthoryear{{Deck}, {Payne}  \& {Holman}}{{Deck}
  et~al.}{2013}]{Deck2013}
{Deck} K.~M.,  {Payne} M.,   {Holman} M.~J.,  2013, \mn@doi [\apj]
  {10.1088/0004-637X/774/2/129}, \href
  {https://ui.adsabs.harvard.edu/abs/2013ApJ...774..129D} {774, 129}

\bibitem[\protect\citeauthoryear{{Delisle} \& {Laskar}}{{Delisle} \&
  {Laskar}}{2014}]{Delisle2014}
{Delisle} J.~B.,  {Laskar} J.,  2014, \mn@doi [\aap]
  {10.1051/0004-6361/201424227}, \href
  {https://ui.adsabs.harvard.edu/abs/2014A&A...570L...7D} {570, L7}

\bibitem[\protect\citeauthoryear{{Deng}, {Mayer}  \& {Meru}}{{Deng}
  et~al.}{2017}]{Deng2017}
{Deng} H.,  {Mayer} L.,   {Meru} F.,  2017, \mn@doi [\apj]
  {10.3847/1538-4357/aa872b}, \href
  {https://ui.adsabs.harvard.edu/abs/2017ApJ...847...43D} {847, 43}

\bibitem[\protect\citeauthoryear{{Deng}, {Mayer}  \& {Latter}}{{Deng}
  et~al.}{2020}]{deng2020}
{Deng} H.,  {Mayer} L.,   {Latter} H.,  2020, \mn@doi [\apj]
  {10.3847/1538-4357/ab77b2}, \href
  {https://ui.adsabs.harvard.edu/abs/2020ApJ...891..154D} {891, 154}

\bibitem[\protect\citeauthoryear{{Desch} \& {Turner}}{{Desch} \&
  {Turner}}{2015}]{Desch2015}
{Desch} S.~J.,  {Turner} N.~J.,  2015, \mn@doi [\apj]
  {10.1088/0004-637X/811/2/156}, \href
  {https://ui.adsabs.harvard.edu/abs/2015ApJ...811..156D} {811, 156}

\bibitem[\protect\citeauthoryear{{Fabrycky} et~al.,}{{Fabrycky}
  et~al.}{2014}]{Fabrycky2014}
{Fabrycky} D.~C.,  et~al., 2014, \mn@doi [\apj] {10.1088/0004-637X/790/2/146},
  \href {https://ui.adsabs.harvard.edu/abs/2014ApJ...790..146F} {790, 146}

\bibitem[\protect\citeauthoryear{{Flaherty} et~al.,}{{Flaherty}
  et~al.}{2020}]{Flaherty2020}
{Flaherty} K.,  et~al., 2020, \mn@doi [\apj] {10.3847/1538-4357/ab8cc5}, \href
  {https://ui.adsabs.harvard.edu/abs/2020ApJ...895..109F} {895, 109}

\bibitem[\protect\citeauthoryear{{Gammie}}{{Gammie}}{1996}]{Gammie1996}
{Gammie} C.~F.,  1996, \mn@doi [\apj] {10.1086/176735}, \href
  {https://ui.adsabs.harvard.edu/abs/1996ApJ...457..355G} {457, 355}

\bibitem[\protect\citeauthoryear{{Garaud} \& {Lin}}{{Garaud} \&
  {Lin}}{2007}]{GaraudLin2007}
{Garaud} P.,  {Lin} D.~N.~C.,  2007, \mn@doi [\apj] {10.1086/509041}, \href
  {https://ui.adsabs.harvard.edu/abs/2007ApJ...654..606G} {654, 606}

\bibitem[\protect\citeauthoryear{{Goldreich} \& {Schlichting}}{{Goldreich} \&
  {Schlichting}}{2014}]{Goldreich-Schlichting2014}
{Goldreich} P.,  {Schlichting} H.~E.,  2014, \mn@doi [\aj]
  {10.1088/0004-6256/147/2/32}, \href
  {https://ui.adsabs.harvard.edu/abs/2014AJ....147...32G} {147, 32}

\bibitem[\protect\citeauthoryear{{Goldreich} \& {Tremaine}}{{Goldreich} \&
  {Tremaine}}{1980}]{Goldreich1980}
{Goldreich} P.,  {Tremaine} S.,  1980, \mn@doi [\apj] {10.1086/158356}, \href
  {https://ui.adsabs.harvard.edu/abs/1980ApJ...241..425G} {241, 425}

\bibitem[\protect\citeauthoryear{{Guilera}, {Benitez-Llambay}, {Miller
  Bertolami}  \& {Pessah}}{{Guilera} et~al.}{2023}]{Guilera2023}
{Guilera} O.~M.,  {Benitez-Llambay} P.,  {Miller Bertolami} M.~M.,   {Pessah}
  M.~E.,  2023, \mn@doi [\apj] {10.3847/1538-4357/acd2cb}, \href
  {https://ui.adsabs.harvard.edu/abs/2023ApJ...953...97G} {953, 97}

\bibitem[\protect\citeauthoryear{{Hands} \& {Alexander}}{{Hands} \&
  {Alexander}}{2018}]{Hands-Alexander2018}
{Hands} T.~O.,  {Alexander} R.~D.,  2018, \mn@doi [\mnras]
  {10.1093/mnras/stx2711}, \href
  {https://ui.adsabs.harvard.edu/abs/2018MNRAS.474.3998H} {474, 3998}

\bibitem[\protect\citeauthoryear{{Hands}, {Alexander}  \& {Dehnen}}{{Hands}
  et~al.}{2014}]{Hands2014}
{Hands} T.~O.,  {Alexander} R.~D.,   {Dehnen} W.,  2014, \mn@doi [\mnras]
  {10.1093/mnras/stu1751}, \href
  {https://ui.adsabs.harvard.edu/abs/2014MNRAS.445..749H} {445, 749}

\bibitem[\protect\citeauthoryear{{Hartmann}, {Calvet}, {Gullbring}  \&
  {D'Alessio}}{{Hartmann} et~al.}{1998}]{hartmann1998}
{Hartmann} L.,  {Calvet} N.,  {Gullbring} E.,   {D'Alessio} P.,  1998, \mn@doi
  [\apj] {10.1086/305277}, \href
  {https://ui.adsabs.harvard.edu/abs/1998ApJ...495..385H} {495, 385}

\bibitem[\protect\citeauthoryear{{Hou} \& {Yu}}{{Hou} \& {Yu}}{2024}]{Hou2024}
{Hou} Q.,  {Yu} C.,  2024, \mn@doi [\apj] {10.3847/1538-4357/ad6a5c}, \href
  {https://ui.adsabs.harvard.edu/abs/2024ApJ...972..152H} {972, 152}

\bibitem[\protect\citeauthoryear{{Hou} \& {Yu}}{{Hou} \& {Yu}}{2025}]{Hou2025}
{Hou} Q.,  {Yu} C.,  2025, \mn@doi [\apj] {10.3847/1538-4357/ada15a}, \href
  {https://ui.adsabs.harvard.edu/abs/2025ApJ...979..185H} {979, 185}

\bibitem[\protect\citeauthoryear{{Huang} \& {Ormel}}{{Huang} \&
  {Ormel}}{2023}]{HuangOrmel2023}
{Huang} S.,  {Ormel} C.~W.,  2023, \mn@doi [\mnras] {10.1093/mnras/stad1032},
  \href {https://ui.adsabs.harvard.edu/abs/2023MNRAS.522..828H} {522, 828}

\bibitem[\protect\citeauthoryear{{Ida} \& {Lin}}{{Ida} \&
  {Lin}}{2008}]{ida2008}
{Ida} S.,  {Lin} D.~N.~C.,  2008, \mn@doi [\apj] {10.1086/523754}, \href
  {https://ui.adsabs.harvard.edu/abs/2008ApJ...673..487I} {673, 487}

\bibitem[\protect\citeauthoryear{{Ida}, {Muto}, {Matsumura}  \&
  {Brasser}}{{Ida} et~al.}{2020}]{Ida2020}
{Ida} S.,  {Muto} T.,  {Matsumura} S.,   {Brasser} R.,  2020, \mn@doi [\mnras]
  {10.1093/mnras/staa1073}, \href
  {https://ui.adsabs.harvard.edu/abs/2020MNRAS.494.5666I} {494, 5666}

\bibitem[\protect\citeauthoryear{{Izidoro}, {Ogihara}, {Raymond}, {Morbidelli},
  {Pierens}, {Bitsch}, {Cossou}  \& {Hersant}}{{Izidoro}
  et~al.}{2017}]{Izidoro2017}
{Izidoro} A.,  {Ogihara} M.,  {Raymond} S.~N.,  {Morbidelli} A.,  {Pierens} A.,
   {Bitsch} B.,  {Cossou} C.,   {Hersant} F.,  2017, \mn@doi [\mnras]
  {10.1093/mnras/stx1232}, \href
  {https://ui.adsabs.harvard.edu/abs/2017MNRAS.470.1750I} {470, 1750}

\bibitem[\protect\citeauthoryear{{Kanagawa}, {Tanaka}, {Muto}, {Tanigawa}  \&
  {Takeuchi}}{{Kanagawa} et~al.}{2015}]{Kanagawa2015}
{Kanagawa} K.~D.,  {Tanaka} H.,  {Muto} T.,  {Tanigawa} T.,   {Takeuchi} T.,
  2015, \mn@doi [\mnras] {10.1093/mnras/stv025}, \href
  {https://ui.adsabs.harvard.edu/abs/2015MNRAS.448..994K} {448, 994}

\bibitem[\protect\citeauthoryear{{Keller}, {Dai}  \& {Xu}}{{Keller}
  et~al.}{2025}]{Keller2025}
{Keller} F.,  {Dai} F.,   {Xu} W.,  2025, \mn@doi [arXiv e-prints]
  {10.48550/arXiv.2504.12596}, \href
  {https://ui.adsabs.harvard.edu/abs/2025arXiv250412596K} {p. arXiv:2504.12596}

\bibitem[\protect\citeauthoryear{{Kley} \& {Nelson}}{{Kley} \&
  {Nelson}}{2012}]{Kley2012}
{Kley} W.,  {Nelson} R.~P.,  2012, \mn@doi [\araa]
  {10.1146/annurev-astro-081811-125523}, \href
  {https://ui.adsabs.harvard.edu/abs/2012ARA&A..50..211K} {50, 211}

\bibitem[\protect\citeauthoryear{{Kubli}, {Mayer}, {Deng}  \& {Lin}}{{Kubli}
  et~al.}{2025}]{Kubli2025}
{Kubli} N.,  {Mayer} L.,  {Deng} H.,   {Lin} D. N.~C.,  2025, \mn@doi [arXiv
  e-prints] {10.48550/arXiv.2503.01973}, \href
  {https://ui.adsabs.harvard.edu/abs/2025arXiv250301973K} {p. arXiv:2503.01973}

\bibitem[\protect\citeauthoryear{{Lammers}, {Hadden}  \& {Murray}}{{Lammers}
  et~al.}{2024}]{Lammers2024}
{Lammers} C.,  {Hadden} S.,   {Murray} N.,  2024, \mn@doi [\apj]
  {10.3847/1538-4357/ad5be6}, \href
  {https://ui.adsabs.harvard.edu/abs/2024ApJ...972...53L} {972, 53}

\bibitem[\protect\citeauthoryear{{Laughlin}, {Steinacker}  \&
  {Adams}}{{Laughlin} et~al.}{2004}]{Laughlin2004}
{Laughlin} G.,  {Steinacker} A.,   {Adams} F.~C.,  2004, \mn@doi [\apj]
  {10.1086/386316}, \href
  {https://ui.adsabs.harvard.edu/abs/2004ApJ...608..489L} {608, 489}

\bibitem[\protect\citeauthoryear{{Lee} \& {Peale}}{{Lee} \&
  {Peale}}{2002}]{LeePeale2002}
{Lee} M.~H.,  {Peale} S.~J.,  2002, \mn@doi [\apj] {10.1086/338504}, \href
  {https://ui.adsabs.harvard.edu/abs/2002ApJ...567..596L} {567, 596}

\bibitem[\protect\citeauthoryear{{Lee}, {Fabrycky}  \& {Lin}}{{Lee}
  et~al.}{2013}]{Lee2013}
{Lee} M.~H.,  {Fabrycky} D.,   {Lin} D.~N.~C.,  2013, \mn@doi [\apj]
  {10.1088/0004-637X/774/1/52}, \href
  {https://ui.adsabs.harvard.edu/abs/2013ApJ...774...52L} {774, 52}

\bibitem[\protect\citeauthoryear{{Lesur} et~al.,}{{Lesur}
  et~al.}{2023}]{Lesur-etal.2023-PPVII}
{Lesur} G.,  et~al., 2023, in {Inutsuka} S.,  {Aikawa} Y.,  {Muto} T.,
  {Tomida} K.,   {Tamura} M.,  eds,  Astronomical Society of the Pacific
  Conference Series Vol. 534, Protostars and Planets VII. p.~465 (\mn@eprint
  {arXiv} {2203.09821}), \mn@doi{10.48550/arXiv.2203.09821}

\bibitem[\protect\citeauthoryear{{Li}, {Li}, {Li}  \& {Lin}}{{Li}
  et~al.}{2019}]{Li2019}
{Li} Y.-P.,  {Li} H.,  {Li} S.,   {Lin} D. N.~C.,  2019, \mn@doi [\apj]
  {10.3847/1538-4357/ab4bc8}, \href
  {https://ui.adsabs.harvard.edu/abs/2019ApJ...886...62L} {886, 62}

\bibitem[\protect\citeauthoryear{{Li}, {Chiang}, {Choksi}  \& {Dai}}{{Li}
  et~al.}{2024}]{RixinLi2024}
{Li} R.,  {Chiang} E.,  {Choksi} N.,   {Dai} F.,  2024, \mn@doi [arXiv
  e-prints] {10.48550/arXiv.2408.10206}, \href
  {https://ui.adsabs.harvard.edu/abs/2024arXiv240810206L} {p. arXiv:2408.10206}

\bibitem[\protect\citeauthoryear{{Lin} \& {Papaloizou}}{{Lin} \&
  {Papaloizou}}{1979}]{linpap1979}
{Lin} D.~N.~C.,  {Papaloizou} J.,  1979, \mn@doi [\mnras]
  {10.1093/mnras/188.2.191}, \href
  {https://ui.adsabs.harvard.edu/abs/1979MNRAS.188..191L} {188, 191}

\bibitem[\protect\citeauthoryear{{Lin} \& {Papaloizou}}{{Lin} \&
  {Papaloizou}}{1986}]{LinPapaloizou1986}
{Lin} D.~N.~C.,  {Papaloizou} J.,  1986, \mn@doi [\apj] {10.1086/164653}, \href
  {https://ui.adsabs.harvard.edu/abs/1986ApJ...309..846L} {309, 846}

\bibitem[\protect\citeauthoryear{{Lin}, {Liu}  \& {Zheng}}{{Lin}
  et~al.}{2025}]{Lin2025}
{Lin} L.,  {Liu} B.,   {Zheng} Z.,  2025, \mn@doi [arXiv e-prints]
  {10.48550/arXiv.2501.12650}, \href
  {https://ui.adsabs.harvard.edu/abs/2025arXiv250112650L} {p. arXiv:2501.12650}

\bibitem[\protect\citeauthoryear{{Lithwick} \& {Wu}}{{Lithwick} \&
  {Wu}}{2012}]{LithwickWu2012}
{Lithwick} Y.,  {Wu} Y.,  2012, \mn@doi [\apjl] {10.1088/2041-8205/756/1/L11},
  \href {https://ui.adsabs.harvard.edu/abs/2012ApJ...756L..11L} {756, L11}

\bibitem[\protect\citeauthoryear{{Liu}, {Zhang}, {Lin}  \& {Aarseth}}{{Liu}
  et~al.}{2015}]{liu2015}
{Liu} B.,  {Zhang} X.,  {Lin} D. N.~C.,   {Aarseth} S.~J.,  2015, \mn@doi
  [\apj] {10.1088/0004-637X/798/1/62}, \href
  {https://ui.adsabs.harvard.edu/abs/2015ApJ...798...62L} {798, 62}

\bibitem[\protect\citeauthoryear{{Masset}}{{Masset}}{2000}]{Masset2000}
{Masset} F.,  2000, \mn@doi [\aaps] {10.1051/aas:2000116}, \href
  {https://ui.adsabs.harvard.edu/abs/2000A&AS..141..165M} {141, 165}

\bibitem[\protect\citeauthoryear{{McNally}, {Nelson}, {Paardekooper}, {Gressel}
   \& {Lyra}}{{McNally} et~al.}{2017}]{McNally2017}
{McNally} C.~P.,  {Nelson} R.~P.,  {Paardekooper} S.-J.,  {Gressel} O.,
  {Lyra} W.,  2017, \mn@doi [\mnras] {10.1093/mnras/stx2136}, \href
  {https://ui.adsabs.harvard.edu/abs/2017MNRAS.472.1565M} {472, 1565}

\bibitem[\protect\citeauthoryear{{McNally}, {Nelson}  \&
  {Paardekooper}}{{McNally} et~al.}{2018}]{McNally2018}
{McNally} C.~P.,  {Nelson} R.~P.,   {Paardekooper} S.-J.,  2018, \mn@doi
  [\mnras] {10.1093/mnras/sty905}, \href
  {https://ui.adsabs.harvard.edu/abs/2018MNRAS.477.4596M} {477, 4596}

\bibitem[\protect\citeauthoryear{{Meyer} \& {Wisdom}}{{Meyer} \&
  {Wisdom}}{2008}]{MeyerWisdom2008}
{Meyer} J.,  {Wisdom} J.,  2008, \mn@doi [\icarus]
  {10.1016/j.icarus.2007.09.008}, \href
  {https://ui.adsabs.harvard.edu/abs/2008Icar..193..213M} {193, 213}

\bibitem[\protect\citeauthoryear{{Murray} \& {Dermott}}{{Murray} \&
  {Dermott}}{1999}]{Murray-Dermott1999}
{Murray} C.~D.,  {Dermott} S.~F.,  1999, {Solar System Dynamics},
  \mn@doi{10.1017/CBO9781139174817.
}

\bibitem[\protect\citeauthoryear{{Nayakshin}, {Cruz S{\'a}enz de Miera},
  {K{\'o}sp{\'a}l}, {{\'C}alovi{\'c}}, {Eisl{\"o}ffel}  \& {Lin}}{{Nayakshin}
  et~al.}{2024}]{Nayakshin-etal.2024}
{Nayakshin} S.,  {Cruz S{\'a}enz de Miera} F.,  {K{\'o}sp{\'a}l} {\'A}.,
  {{\'C}alovi{\'c}} A.,  {Eisl{\"o}ffel} J.,   {Lin} D. N.~C.,  2024, \mn@doi
  [\mnras] {10.1093/mnras/stae877}, \href
  {https://ui.adsabs.harvard.edu/abs/2024MNRAS.530.1749N} {530, 1749}

\bibitem[\protect\citeauthoryear{{Nelson}}{{Nelson}}{2005}]{Nelson2005}
{Nelson} R.~P.,  2005, \mn@doi [\aap] {10.1051/0004-6361:20042605}, \href
  {https://ui.adsabs.harvard.edu/abs/2005A&A...443.1067N} {443, 1067}

\bibitem[\protect\citeauthoryear{{Ogihara}, {Ida}  \& {Morbidelli}}{{Ogihara}
  et~al.}{2007}]{Ogihara2007}
{Ogihara} M.,  {Ida} S.,   {Morbidelli} A.,  2007, \mn@doi [\icarus]
  {10.1016/j.icarus.2006.12.006}, \href
  {https://ui.adsabs.harvard.edu/abs/2007Icar..188..522O} {188, 522}

\bibitem[\protect\citeauthoryear{{Paardekooper}, {Baruteau}, {Crida}  \&
  {Kley}}{{Paardekooper} et~al.}{2010}]{Paardekooper2010}
{Paardekooper} S.~J.,  {Baruteau} C.,  {Crida} A.,   {Kley} W.,  2010, \mn@doi
  [\mnras] {10.1111/j.1365-2966.2009.15782.x}, \href
  {https://ui.adsabs.harvard.edu/abs/2010MNRAS.401.1950P} {401, 1950}

\bibitem[\protect\citeauthoryear{{Paardekooper}, {Rein}  \&
  {Kley}}{{Paardekooper} et~al.}{2013}]{Paardekooper2013}
{Paardekooper} S.-J.,  {Rein} H.,   {Kley} W.,  2013, \mn@doi [\mnras]
  {10.1093/mnras/stt1224}, \href
  {https://ui.adsabs.harvard.edu/abs/2013MNRAS.434.3018P} {434, 3018}

\bibitem[\protect\citeauthoryear{{Paardekooper}, {Dong}, {Duffell}, {Fung},
  {Masset}, {Ogilvie}  \& {Tanaka}}{{Paardekooper}
  et~al.}{2023}]{Paardekooper2023}
{Paardekooper} S.,  {Dong} R.,  {Duffell} P.,  {Fung} J.,  {Masset} F.~S.,
  {Ogilvie} G.,   {Tanaka} H.,  2023, in {Inutsuka} S.,  {Aikawa} Y.,  {Muto}
  T.,  {Tomida} K.,   {Tamura} M.,  eds,  Astronomical Society of the Pacific
  Conference Series Vol. 534, Protostars and Planets VII. p.~685 (\mn@eprint
  {arXiv} {2203.09595}), \mn@doi{10.48550/arXiv.2203.09595}

\bibitem[\protect\citeauthoryear{{Papaloizou} \& {Larwood}}{{Papaloizou} \&
  {Larwood}}{2000}]{Papaloizou2000}
{Papaloizou} J.~C.~B.,  {Larwood} J.~D.,  2000, \mn@doi [\mnras]
  {10.1046/j.1365-8711.2000.03466.x}, \href
  {https://ui.adsabs.harvard.edu/abs/2000MNRAS.315..823P} {315, 823}

\bibitem[\protect\citeauthoryear{{Papaloizou} \& {Terquem}}{{Papaloizou} \&
  {Terquem}}{2010}]{PapaloizouTerquem2010}
{Papaloizou} J. C.~B.,  {Terquem} C.,  2010, \mn@doi [\mnras]
  {10.1111/j.1365-2966.2010.16477.x}, \href
  {https://ui.adsabs.harvard.edu/abs/2010MNRAS.405..573P} {405, 573}

\bibitem[\protect\citeauthoryear{{Peale}, {Cassen}  \& {Reynolds}}{{Peale}
  et~al.}{1979}]{peale1979}
{Peale} S.~J.,  {Cassen} P.,   {Reynolds} R.~T.,  1979, \mn@doi [Science]
  {10.1126/science.203.4383.892}, \href
  {https://ui.adsabs.harvard.edu/abs/1979Sci...203..892P} {203, 892}

\bibitem[\protect\citeauthoryear{{Rea}, {Simon}, {Carrera}, {Lesur}, {Lyra},
  {Sengupta}, {Yang}  \& {Youdin}}{{Rea} et~al.}{2024}]{Rea-etal.2024}
{Rea} D.~G.,  {Simon} J.~B.,  {Carrera} D.,  {Lesur} G.,  {Lyra} W.,
  {Sengupta} D.,  {Yang} C.-C.,   {Youdin} A.~N.,  2024, \mn@doi [\apj]
  {10.3847/1538-4357/ad57c5}, \href
  {https://ui.adsabs.harvard.edu/abs/2024ApJ...972..128R} {972, 128}

\bibitem[\protect\citeauthoryear{{Rein} \& {Papaloizou}}{{Rein} \&
  {Papaloizou}}{2009}]{Rein-Papaloizou2009}
{Rein} H.,  {Papaloizou} J.~C.~B.,  2009, \mn@doi [\aap]
  {10.1051/0004-6361/200811330}, \href
  {https://ui.adsabs.harvard.edu/abs/2009A&A...497..595R} {497, 595}

\bibitem[\protect\citeauthoryear{{Ricci}, {Liu}, {Isella}  \& {Li}}{{Ricci}
  et~al.}{2018}]{Ricci-etal.2018}
{Ricci} L.,  {Liu} S.-F.,  {Isella} A.,   {Li} H.,  2018, \mn@doi [\apj]
  {10.3847/1538-4357/aaa546}, \href
  {https://ui.adsabs.harvard.edu/abs/2018ApJ...853..110R} {853, 110}

\bibitem[\protect\citeauthoryear{{Rice}, {Armitage}, {Bate}  \&
  {Bonnell}}{{Rice} et~al.}{2003}]{Rice2003}
{Rice} W.~K.~M.,  {Armitage} P.~J.,  {Bate} M.~R.,   {Bonnell} I.~A.,  2003,
  \mn@doi [\mnras] {10.1046/j.1365-8711.2003.06253.x}, \href
  {https://ui.adsabs.harvard.edu/abs/2003MNRAS.339.1025R} {339, 1025}

\bibitem[\protect\citeauthoryear{{Romero-Mirza} et~al.,}{{Romero-Mirza}
  et~al.}{2024}]{Romero-Mirza2024}
{Romero-Mirza} C.~E.,  et~al., 2024, \mn@doi [\apj] {10.3847/1538-4357/ad20e9},
  \href {https://ui.adsabs.harvard.edu/abs/2024ApJ...964...36R} {964, 36}

\bibitem[\protect\citeauthoryear{{Rosotti}}{{Rosotti}}{2023}]{Rosotti2023}
{Rosotti} G.~P.,  2023, \mn@doi [\nar] {10.1016/j.newar.2023.101674}, \href
  {https://ui.adsabs.harvard.edu/abs/2023NewAR..9601674R} {96, 101674}

\bibitem[\protect\citeauthoryear{{Salyk}, {Pontoppidan}, {Blake}, {Lahuis},
  {van Dishoeck}  \& {Evans}}{{Salyk} et~al.}{2008}]{Salyk2008}
{Salyk} C.,  {Pontoppidan} K.~M.,  {Blake} G.~A.,  {Lahuis} F.,  {van Dishoeck}
  E.~F.,   {Evans} II N.~J.,  2008, \mn@doi [\apjl] {10.1086/586894}, \href
  {https://ui.adsabs.harvard.edu/abs/2008ApJ...676L..49S} {676, L49}

\bibitem[\protect\citeauthoryear{{Secunda}, {Bellovary}, {Mac Low}, {Ford},
  {McKernan}, {Leigh}, {Lyra}  \& {S{\'a}ndor}}{{Secunda}
  et~al.}{2019}]{Secunda2019}
{Secunda} A.,  {Bellovary} J.,  {Mac Low} M.-M.,  {Ford} K.~E.~S.,  {McKernan}
  B.,  {Leigh} N. W.~C.,  {Lyra} W.,   {S{\'a}ndor} Z.,  2019, \mn@doi [\apj]
  {10.3847/1538-4357/ab20ca}, \href
  {https://ui.adsabs.harvard.edu/abs/2019ApJ...878...85S} {878, 85}

\bibitem[\protect\citeauthoryear{{Shakura} \& {Sunyaev}}{{Shakura} \&
  {Sunyaev}}{1973}]{shakura1973}
{Shakura} N.~I.,  {Sunyaev} R.~A.,  1973, \aap, \href
  {https://ui.adsabs.harvard.edu/abs/1973A&A....24..337S} {24, 337}

\bibitem[\protect\citeauthoryear{{Simon}, {Beckwith}  \& {Armitage}}{{Simon}
  et~al.}{2012}]{Simon2012}
{Simon} J.~B.,  {Beckwith} K.,   {Armitage} P.~J.,  2012, \mn@doi [\mnras]
  {10.1111/j.1365-2966.2012.20835.x}, \href
  {https://ui.adsabs.harvard.edu/abs/2012MNRAS.422.2685S} {422, 2685}

\bibitem[\protect\citeauthoryear{{Tanaka}, {Takeuchi}  \& {Ward}}{{Tanaka}
  et~al.}{2002}]{Tanaka2002}
{Tanaka} H.,  {Takeuchi} T.,   {Ward} W.~R.,  2002, \mn@doi [\apj]
  {10.1086/324713}, \href
  {https://ui.adsabs.harvard.edu/abs/2002ApJ...565.1257T} {565, 1257}

\bibitem[\protect\citeauthoryear{{Terquem} \& {Papaloizou}}{{Terquem} \&
  {Papaloizou}}{2019}]{Terquem2019}
{Terquem} C.,  {Papaloizou} J. C.~B.,  2019, \mn@doi [\mnras]
  {10.1093/mnras/sty2693}, \href
  {https://ui.adsabs.harvard.edu/abs/2019MNRAS.482..530T} {482, 530}

\bibitem[\protect\citeauthoryear{{Ward}}{{Ward}}{1997}]{Ward1997}
{Ward} W.~R.,  1997, \mn@doi [\icarus] {10.1006/icar.1996.5647}, \href
  {https://ui.adsabs.harvard.edu/abs/1997Icar..126..261W} {126, 261}

\bibitem[\protect\citeauthoryear{{Wong} \& {Lee}}{{Wong} \&
  {Lee}}{2024}]{Wong2024}
{Wong} K.~H.,  {Lee} M.~H.,  2024, \mn@doi [\aj] {10.3847/1538-3881/ad1f60},
  \href {https://ui.adsabs.harvard.edu/abs/2024AJ....167..112W} {167, 112}

\bibitem[\protect\citeauthoryear{{Wu} \& {Chen}}{{Wu} \&
  {Chen}}{2025}]{Wu-Chen2025}
{Wu} Y.,  {Chen} Y.-X.,  2025, \mn@doi [\mnras] {10.1093/mnrasl/slae102}, \href
  {https://ui.adsabs.harvard.edu/abs/2025MNRAS.536L..13W} {536, L13}

\bibitem[\protect\citeauthoryear{{Wu}, {Baruteau}  \& {Nayakshin}}{{Wu}
  et~al.}{2023}]{Wu_2023}
{Wu} Y.,  {Baruteau} C.,   {Nayakshin} S.,  2023, \mn@doi [\mnras]
  {10.1093/mnras/stad1791}, \href
  {https://ui.adsabs.harvard.edu/abs/2023MNRAS.523.4869W} {523, 4869}

\bibitem[\protect\citeauthoryear{{Wu}, {Chen}  \& {Lin}}{{Wu}
  et~al.}{2024a}]{Wu2024chaotic}
{Wu} Y.,  {Chen} Y.-X.,   {Lin} D. N.~C.,  2024a, \mn@doi [\mnras]
  {10.1093/mnrasl/slad183}, \href
  {https://ui.adsabs.harvard.edu/abs/2024MNRAS.528L.127W} {528, L127}

\bibitem[\protect\citeauthoryear{{Wu}, {Lin}, {Cui}, {Krapp}, {Lee}  \&
  {Youdin}}{{Wu} et~al.}{2024b}]{Wu2024BDHI}
{Wu} Y.,  {Lin} M.-K.,  {Cui} C.,  {Krapp} L.,  {Lee} Y.-N.,   {Youdin} A.~N.,
  2024b, \mn@doi [\apj] {10.3847/1538-4357/ad15fe}, \href
  {https://ui.adsabs.harvard.edu/abs/2024ApJ...962..173W} {962, 173}

\bibitem[\protect\citeauthoryear{{Wu}, {Liu}, {Jiang}  \& {Nayakshin}}{{Wu}
  et~al.}{2024c}]{Wu-etal.2024-SKA}
{Wu} Y.,  {Liu} S.-F.,  {Jiang} H.,   {Nayakshin} S.,  2024c, \mn@doi [\apj]
  {10.3847/1538-4357/ad323b}, \href
  {https://ui.adsabs.harvard.edu/abs/2024ApJ...965..110W} {965, 110}

\bibitem[\protect\citeauthoryear{{Xu} \& {Dai}}{{Xu} \&
  {Dai}}{2025}]{Xu-Dai2025}
{Xu} Y.,  {Dai} F.,  2025, \mn@doi [\apj] {10.3847/1538-4357/adb281}, \href
  {https://ui.adsabs.harvard.edu/abs/2025ApJ...981..142X} {981, 142}

\bibitem[\protect\citeauthoryear{{Yang} \& {Li}}{{Yang} \&
  {Li}}{2024}]{Yangli2024}
{Yang} H.,  {Li} Y.-P.,  2024, \mn@doi [\mnras] {10.1093/mnras/stae2097}, \href
  {https://ui.adsabs.harvard.edu/abs/2024MNRAS.534..485Y} {534, 485}

\bibitem[\protect\citeauthoryear{{de Val-Borro} et~al.,}{{de Val-Borro}
  et~al.}{2006}]{deValBorro2006}
{de Val-Borro} M.,  et~al., 2006, \mn@doi [\mnras]
  {10.1111/j.1365-2966.2006.10488.x}, \href
  {https://ui.adsabs.harvard.edu/abs/2006MNRAS.370..529D} {370, 529}

\makeatother
\end{thebibliography}



\bsp	
\label{lastpage}
\end{CJK*}
\end{document}